\documentclass[twocolumn,preprintnumbers,amsmath,amssymb,aps,prb]{revtex4-1}
\usepackage{bm}
\usepackage{times}
\usepackage{amsmath}
\usepackage{mathrsfs}
\usepackage{graphicx}
\usepackage{epstopdf}
\usepackage[colorlinks=true, letterpaper=ture, pdfstartview=FitV, linkcolor=blue, citecolor=blue, urlcolor=blue]{hyperref}%ÊµÏÖÒýÓÃ¼°Ä¿Â¼Ìø×ª
\usepackage{dcolumn}% Align table columns on decimal point ¶ÔÆëÐ¡ÊýµãÉÏµÄ±íÁÐ
\begin{document}
	
	\title{Photogalvanic effect and second harmonic generation from radio to infrared region in WTe$_2$ monolayer}
	%Nonlinear photocurrents from radio to infrared region in the WTe$_2$ monolayer: A quantum kinetics study
	\author{Yuan Liu$ ^{1,2} $}
	
	\author{Zhen-Gang Zhu$ ^{1,2,3}$}
	\email{zgzhu@ucas.ac.cn}
	
	\author{Gang Su$^4 $}
	\email{gsu@ucas.ac.cn}
	\affiliation{
		$^{1}$ School of Physical Sciences, University of Chinese Academy of Sciences, Beijing 100049, China. \\
		$^{2}$ School of Electronic, Electrical and Communication Engineering, University of Chinese Academy of Sciences, Beijing 100049, China.\\
		$^{3}$ CAS Center for Excellence in Topological Quantum Computation, University of Chinese Academy of Sciences, Beijing 100049, China.\\
		$^{4}$ Kavli Institute for Theoretical Sciences, University of Chinese Academy of Sciences, Beijing 100190, China.
	}
	\date{\today}
	\begin{abstract}
		\textcolor{red}{Second-order nonlinear optical responses, including photogalvanic effect (PGE) and second harmonic generation (SHG),} are fundamental and important physical phenomena in nonlinear optics and optoelectronics. The PGE and SHG associated with linearly and circularly polarized light are called the linear photogalvanic effect (LPGE), circular photogalvanic effect (CPGE), linear second harmonic generation (LSHG), and circular second harmonic generation (CSHG), respectively. In this work, we use the quantum kinetics under the relaxation time approximation to investigate the dependence of second-order nonlinear optical responses on Fermi level and frequency under different out-of-plane electric fields in $ T_d$-WTe$_2$ monolayer from radio to infrared region. We find that the maximum frequency at which the Berry curvature dipole mechanism for the nonlinear Hall effect plays a major role is about 1 THz. From the aspect of Fermi level, in the radio and microwave regions, the two large peaks of nonlinear conductivities occur when the Fermi level is equal to the energy corresponding to the vicinity of the gap-opening points in the band dispersion. From the aspect of frequency, in the radio region, the LPGE and SHG conductivities maintain a large constant while the CPGE conductivity almost disappears. In the microwave region, the LPGE and SHG conductivities start to decrease gradually with increasing frequency while the CPGE conductivity is large. In the infrared region, the frequency and Fermi level dependence of second-order nonlinear optical responses is complicated. In the 125 THz-300 THz region and in the $y$-direction, the presence of DC current without the disturbance of second harmonic current under circularly polarized light may be useful for the fabrication of new optoelectronic devices. Moreover, we illustrate that when calculating the nonlinear Hall effect or second-order nonlinear optical responses of practical materials, the theories in the clean limit fail and it is necessary to use a theory that takes into account scattering effects (e.g., relaxation time approximation). We also point out that for materials with femtosecond-scale relaxation times and complex energy band structures, the quantum kinetics method is more accurate than the semi-classical Boltzmann equation method. Besides, phenomenological expressions of PGE and SHG are provided. Our study is promising to promote the more accurate calculation of second-order nonlinear optical responses in practical materials.
	\end{abstract}
	\maketitle
	
	%\tableofcontents   %²úÉú¿ÉÒÔµã»÷¾Í×Ô¶¯Ìø×ªµÄÄ¿Â¼£¬ÔÚÍ¶¸åµÄÂÛÎÄÖÐ×¢ÊÍÕâÐÐ¾ÍÃ»ÓÐÄ¿Â¼ÁË
	%\newpage           %½«Ä¿Â¼µ¥¶ÀÒ»Ò³
	
	%%%%%%%%%%%%%%%%%%%%%%%%%%%%%%%%%%%%%%%%%%%%%%%%%%%%%%%%%%%%%%%%%%%%%%%%%%%%%%%%%%%%%%%%%%%%%%%%%%%%%%%%%%%%%%%%%%%%%%%%%%%%%%%%
	\section{Introduction}
	\par \textcolor{red}{Nonlinear optical phenomena in solids} can be used to probe symmetry breaking, new phases of materials as well as quantum geometry and topology \cite{Ma2021}. Second-order nonlinear optical responses under monochromatic light can be classified as the photogalvanic effect (PGE) and the second harmonic generation (SHG). The PGE, also known as the bulk photovoltaic effect, refers to the generation of DC current when light strikes a homogeneous material that lacks inversion symmetry \cite{Sturman1992,Ganichev2006,Xu2021}. The PGE associated with linearly and circularly polarized light are called linear photogalvanic effect (LPGE) and circular photogalvanic effect (CPGE), respectively \cite{Sturman1992,Ganichev2006}.
	The SHG refers to the generation of \textcolor{red}{frequency-doubled current and the second harmonic radiated from it} when light or an alternating electric field is applied to a homogeneous material that lacks inversion symmetry \cite{He2021,Patankar2018,Ma2018}. The SHG associated with linearly and circularly polarized light\cite{Seyler2015,Hsieh2010} are called linear second harmonic generation (LSHG) and circular second harmonic generation (CSHG), respectively.
	\par The nonlinear Hall effect (NHE) is similar to the low-frequency version of LPGE and LSHG, but it only considers the current transverse to the alternating electric field \cite{Sodemann2015,Ma2021}. The semi-classical Boltzmann equation method reveals that this transverse current originates from the Berry curvature dipole (BCD) \cite{Sodemann2015}, which has been verified by many experiments \cite{Ma2018,Xiao2020,Qin2021}. Recent studies have found that BCD can also be reproduced in quantum kinetics, which also reveals the existence of injection, shift and rectification terms in the PGE besides BCD term \cite{Matsyshyn2019,Watanabe2021,Lihm2022}. Hence it is worth exploring the role of these terms in nonlinear transport for practical materials.

	Recently, the WTe$_2$ monolayer has attracted a lot of attention from experimental and theoretical aspects due to its exotic properties such as quantum spin Hall states \cite{Shi2019a,Zhao2021}, superconductivity \cite{Luepke2020}, NHE \cite{You2018}, CPGE \cite{Xu2018} and SHG \cite{Bhalla2022a,Bhalla2022}. These studies imply the non-trivial geometrical nature of energy band in the WTe$_2$ monolayer.

\begin{figure}[t!]
		\begin{center}
			\includegraphics[width=1\columnwidth]{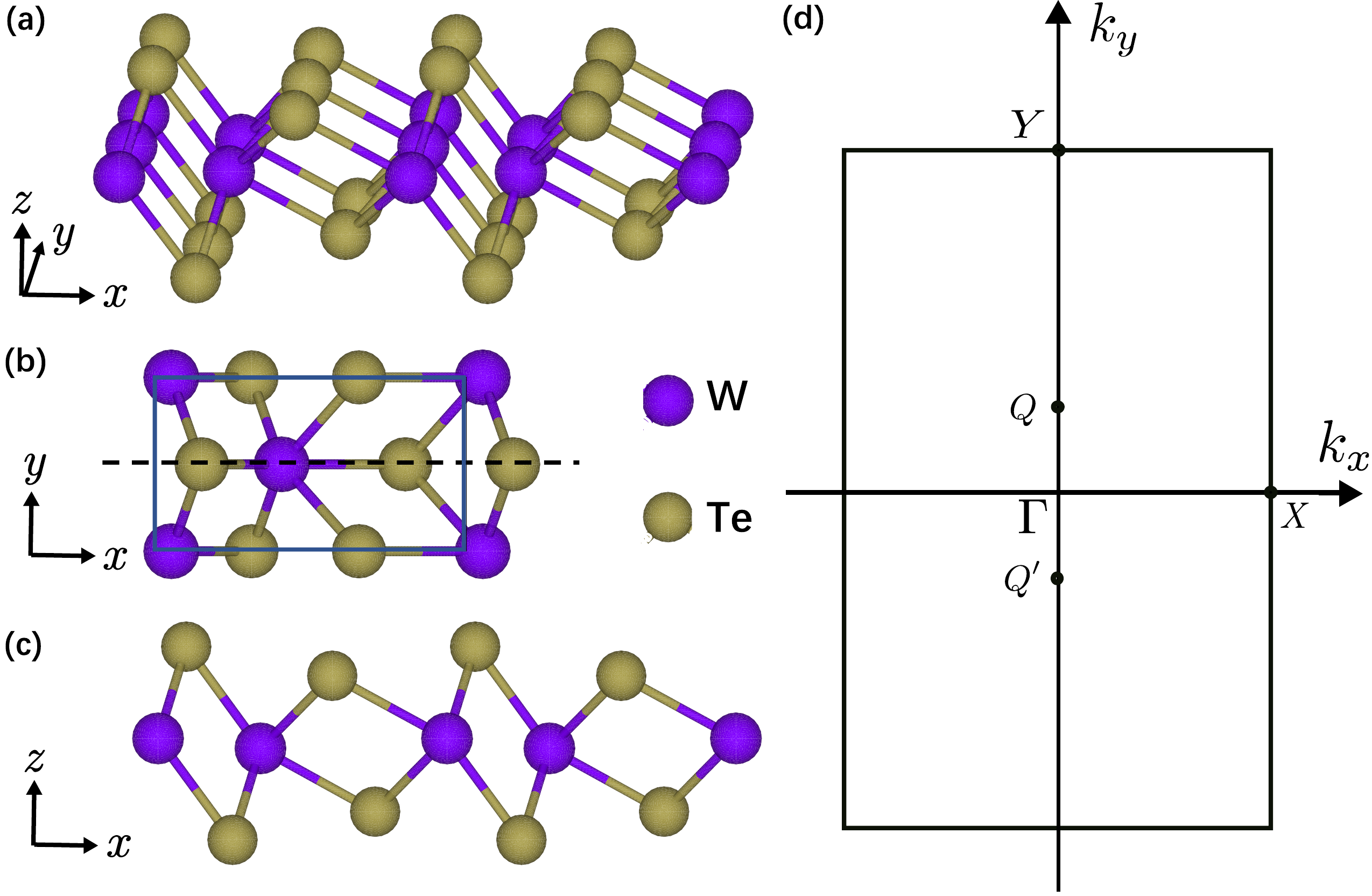}
		\end{center}
		\caption{ (Color online) (a,b,c) The structure of monolayer $ T_d$-WTe$_2$. (d) The first Brillouin zone of $ T_d$-WTe$_2$. The coordinates of $ X$, $Y $, $ Q $ and $ Q^{\prime}$ are $ (0.5, 0) $, $ (0, 0.9) $, $ (0, 0.385) $ and $ (0, -0.385) $ $ \text{\AA}^{-1} $, respectively. $ Q $ and $ Q^{\prime}$ are the gap-opening points \cite{Shi2019}.} \label{Fig1}
	\end{figure}

	The WTe$_2$ monolayer has two phases, the 1$ T^{\prime} $ phase and the $ T_d $ phase, which differ very little \cite{Xu2018} and can be roughly considered as the same phase \cite{Dong2022}. The 1$ T^{\prime} $ phase has perfect inversion symmetry and the $ T_d $ phase weakly breaks the inversion symmetry \cite{Xu2018,Shi2019,Bhalla2022}. The $ T_d$-WTe$_2$ monolayer only has the mirror symmetry $\mathcal{M}_y$ [see the dashed line in Fig.~\ref{Fig1}(b)] and its point group is $ C_{1s} $, which is actually the symmetry existing in the dual-gated experiments \cite{Xu2018}. When a vertical external electric field is applied by dual gates, the inversion symmetry of the WTe$_2$ monolayer is more strongly broken and it is induced to produce a net dipole moment, which strongly affects the in-plane transport properties \cite{Xu2018,Shi2019}.
	
	The experiment of Ref.~\onlinecite{Ma2018} has shown that the NHE of $ T_d$-WTe$_2$ at frequencies below 1000 Hz is contributed by BCD. Therefore, two questions then arise: What is the frequency upper limit below which the BCD mechanism of $ T_d$-WTe$_2$ monolayer can play a major role? What are the main mechanisms of second-order nonlinear optical responses at higher frequencies?
	In addition, the dependence of the CPGE of $ T_d$-WTe$_2$ monolayer on the out-of-plane electric field $ E_{\perp} $ at a fixed Fermi level and frequency of 29 THz has been studied in Ref.~\onlinecite{Xu2018} by mid-infrared optoelectronic microscopy and the theory of injection current in the two-band limit. However, the general dependence of PGE and SHG on frequency and Fermi level remains unclear.

	In this paper, we use the quantum kinetics under the relaxation time approximation \cite{Matsyshyn2019} to investigate the dependence of PGE and SHG on Fermi level and frequency under different $ E_{\perp} $ in $ T_d$-WTe$_2$ monolayer.
	The theory shows that second-order nonlinear conductivities can be classified into Drude, BCD, interband 2 (IB2), interband 3 (IB3) terms \cite{Matsyshyn2019}. When the relaxation time $ \tau $ is taken as 5 ps, the contribution mechanisms of second-order nonlinear optical responses in $ T_d$-WTe$_2$ monolayer in different frequency regions are summarized in Table~\ref{Contribution mechanisms}, which also shows that the maximum frequency at which the BCD mechanism of the nonlinear Hall effect plays a major role is about 1 THz.
	From the aspect of Fermi level, in the radio and microwave regions, the two large peaks of nonlinear conductivities occur when the Fermi level is equal to the energy corresponding to the vicinity of the gap-opening points in the band dispersion. From the aspect of frequency, in the radio region, the LPGE and SHG conductivities maintain a large constant while the CPGE conductivity almost disappears. In the microwave region, the LPGE and SHG conductivities start to decrease gradually with increasing frequency while the CPGE conductivity is large. In the infrared region, the frequency and Fermi level dependence of second-order nonlinear optical responses is complicated. In the 125 THz-300 THz region and in the $y$-direction, the presence of CPGE without the disturbance of CSHG may be useful for the fabrication of new optoelectronic devices (e.g., circularly polarized infrared photodetectors \cite{Zhang2021,Wei2022} and electromagnetic wave energy harvesting rectifiers \cite{Zhou2020}). Moreover, we illustrate that when calculating the NHE or second-order nonlinear optical responses of practical materials, the theories in the clean limit (e.g., Refs.~\onlinecite{Aversa1995,Sipe2000,Watanabe2021,Lihm2022}) fail and it is essential to use a theory that takes into account scattering effects (e.g., relaxation time approximation). We also demonstrate that for materials with femtosecond-scale relaxation times and complex energy band structures, the quantum kinetics method is more accurate than the semi-classical Boltzmann equation method.

	We also give phenomenological expressions for PGE and SHG. In previous works on PGE \cite{Sturman1992,Ganichev2006,Ahn2020,Xu2021,Lihm2022,Sipe2000}, it was only noticed that if the real part of a second-order PGE conductivity expression $ \sigma_{(2)}^{\eta \alpha \beta}(\omega,-\omega) $ is symmetric under the exchange of indices $ \alpha $ and $\beta $, it relates to linearly polarized light; and if its imaginary part is antisymmetric under the exchange of indices $ \alpha $ and $\beta $, it relates to circularly polarized light.
	According to this principle, it can be judged that the shift and injection currents in the clean limit are related to LPGE and CPGE, respectively \cite{Ahn2020,Xu2021,Sipe2000}. In addition, the procedure of transforming the conductivity to a new symmetrized expression Eq.~(\ref{new-2nd-conductivity}) was previously used in the early work Ref.~\onlinecite{Sipe2000} in order to facilitate the simplification and analysis of conductivity expressions. In this work, we find a new application of this symmetrization procedure, i.e., for any PGE conductivity expression, we only need to transform it into Eq.~(\ref{new-2nd-conductivity}) and then take the real and imaginary parts of this new expression to obtain the conductivity formulas of LPGE and CPGE. \textcolor{red}{Unlike PGE, for any SHG conductivity expression, we only need to transform it to Eq.~(\ref{new-2nd-conductivity}) and then take the modulus to obtain the effective LSHG and CSHG conductivities.}
	%Unlike PGE, for SHG, we find that both the real and imaginary parts of the new symmetrized conductivity expression contribute to LSHG and CSHG.
	%\textcolor{red}{obtained from microscopic theory}
	
	\begin{table*}
		\caption{\label{Contribution mechanisms}Contribution mechanisms of PGE and SHG in $ T_d$-WTe$_2$ monolayer in different frequency regions (relaxation time $ \tau = 5 $ ps). The division of the frequency regions refers to Ref.~\onlinecite{cavell2018national}.}
		\begin{ruledtabular}
			\begin{tabular}{cccccc}
				&\multicolumn{1}{c}{radio region}&\multicolumn{1}{c}{microwave region}&\multicolumn{3}{c}{infrared region}\\
				&$ <0.3 \mathrm{~GHz} $&$ 0.3 \mathrm{~GHz}-0.3 \mathrm{~THz} $&$ 0.3 \mathrm{~THz}-50 \mathrm{~THz}$&$ 50 \mathrm{~THz}-100 \mathrm{~THz}$
				&$ 100 \mathrm{~THz}-400 \mathrm{~THz}$\\ \hline
				Contribution to LPGE&BCD&BCD\footnote{For LPGE, the frequency range of its contribution can be extended to 1 THz.} &None&IB3&None \\
				Contribution to CPGE&None&BCD\footnote{For CPGE, the frequency range of its contribution can be extended to 5 THz.}&None&IB2&IB2\\
				Contribution to LSHG&BCD&BCD
				&BCD,~IB3&BCD,~IB3\footnote{For SHG, the frequency range of their contribution can be extended to 125 THz.}&None\\
				Contribution to CSHG&BCD&BCD
				&BCD,~IB3&BCD,~IB3\footnotemark[3]&None\\
			\end{tabular}
		\end{ruledtabular}
	\end{table*}
	
	\section{Quantum kinetics}
	\textcolor{red}{Although Matsyshyn and Sodemann \cite{Matsyshyn2019} have given nonlinear optical conductivities formulas Eqs.~(\ref{2nd-conductivity-Drude})-(\ref{2nd-conductivity-IB3-shift}) by using quantum kinetics under the relaxation time approximation, we briefly review these basic formulas in this section for completeness. In addition, in this section, we deeply analyze the properties of these basic formulas and specify the importance of nonzero relaxation rate [see also Subsection E of Section VIII of Supplementary Material (SM) \cite{supplemental_material}].} Hereafter we set the alternating electric field as
	\begin{equation}
		\begin{aligned}
			&\vec{E}(t)=\sum_{i} \vec{E}\left(\omega_{i}\right) e^{-i \omega_{i} t}\equiv \vec{E}\left(\omega_{i}\right) e^{-i \omega_{i} t},
		\end{aligned}
	\end{equation}
	where $ \omega_{i} $ is the angular frequency of the light and the repeated index $ \omega_{i} $ indicates summation, e.g., for monochromatic polarized light, $ \omega_{i} = \omega (-\omega)$ as $i=1 (2)$ [see Eq.~(\ref{monochromatic polarized light})].
	In the ``independent particle approximation'' \cite{Sipe2000}, we can set the total Hamiltonian of each electron to be $H_{\text{T}}$:
	\begin{equation}
		\begin{aligned}
			\label{time-evolution-rho-mid}
			\hat{H}_{\text{T}}=&\hat{H}_{0}+\hat{H}_{\text {E}}+\hat{U},\\
			\hat{H}_{0}=&\frac{\hat{\vec{p}}^{2}}{2 m}+{V}(\vec{r}),\\
			\hat{H}_{\text{E}}=&e \hat{\vec{r}} \cdot \vec{E}(t) ,\\
		\end{aligned}
	\end{equation}
	where $ {V}(\vec{r}) $ is the lattice periodic potential, $e>0$ is set and $-e$ is for the charge of electron, $ U $ is the scattering potential, $\hat{H}_{\text{E}}$ represents the electric potential energy of the electron caused by the alternating electric field.
	
	%	\par The time evolution equation of the single particle density operator is [see Supplementary Material (SM) \cite{supplemental_material}]
	%	\begin{equation}\label{time-evolution-rho}
	%		i \hbar \frac{d \hat{\rho}}{d t}=[\hat{H}, \hat{\rho}]=[\hat{H}_{0}+\hat{H}_{\text{E}}+\hat{U}, \hat{\rho}].
	%	\end{equation}	
	%Eq.~(\ref{time-evolution-rho}) becomes
	\par \textcolor{red}{In the relaxation time approximation, the time evolution equation of the single particle density operator is\cite{Matsyshyn2019,Culcer2017,Xu2021,Passos2018} (see SM \cite{supplemental_material})}
	\begin{equation}\label{time-evolution-rho-relax}
		i \hbar \frac{d \hat{\rho}}{d t}-[\hat{H}_{0}+\hat{H}_{\text{E}}, \hat{\rho}]=-i \hbar \Gamma(\hat{\rho}(t)-\hat{\rho}^e),
	\end{equation}
	where the relaxation rate $ \Gamma\equiv {1}/{\tau} $, $ \tau $ is the relaxation time, $ \hat{\rho}^e $ is the single-particle density operator at thermodynamic equilibrium. Due to the complexity of the relaxation processes \cite{Xu2021} and the existence of intraband and interband mixing effects, here we use a uniform $ \tau $ for simplicity, which is a qualitative treatment \cite{Xu2021}. We set the Bloch eigenstate of $\hat{H}_{0}$ as $|n\vec{k}\rangle$, the eigenenergy as $ \varepsilon_{n}(\vec{k}) $ and the periodic part of the Bloch wave $|n\vec{k}\rangle$ as $u_{n \vec{k}}(\vec{r})$. $n$ and $\vec{k}$ are the band index and crystal momentum, respectively.
	We let $f_n(\vec{k})$ be the Fermi-Dirac distribution function corresponding to the state $|n \vec{k}\rangle$, $ f_{mn}(\vec{k})\equiv f_{m}(\vec{k})-f_{n}(\vec{k}) $, $\varepsilon_{m n}(\vec{k}) \equiv \varepsilon_m(\vec{k})-\varepsilon_n(\vec{k})$. The non-Abelian Berry connection is defined as\cite{Matsyshyn2019,Karplus1954}:
	\begin{equation}
		\vec{\xi}_{n m}(\vec{k})=i \int_{N \Omega} u_{n \vec{k} }^{*} \frac{\partial}{\partial \vec{k}} u_{m \vec{k} } d^{3} r,
	\end{equation}
	where $ N\Omega $ is the total crystal volume.
	Hereafter $ \alpha,\beta,\eta $ refer to Cartesian component.

	\par \textcolor{red}{By iterating we can perturbatively solve Eq.~(\ref{time-evolution-rho-relax}) and obtain the first-order density matrix $ {\rho}_{n  m  }^{(1)}(t) $ and the second-order density matrix $ {\rho}_{n  m  }^{(2)}(t) $ (see SM \cite{supplemental_material}).
		The charge current density can be derived by multiplying $ {\rho}_{n  m  }(t) $ with the velocity operator matrix element $v_{m n}^\eta(\vec{k})$ (see SM \cite{supplemental_material}). The discussion of the first-order current density can be seen in SM \cite{supplemental_material}. Next we focus on the second-order current density $ \left\langle j_{\eta}\right\rangle^{(2)} $.}
	Considering that $ \tau $ in the experiments is on the order of picoseconds \cite{Zheng2016} or femtoseconds \cite{Aoki2022,Ma2018,Xu2018,Qin2021}, we can set
	\begin{equation} \begin{aligned} \label{j-2-define}
			\left\langle j_{\eta}\right\rangle^{(2)}\equiv \sigma_{(2)}^{\eta \alpha \beta}\left(\omega_{i}, \omega_{j}\right) E_{\alpha}\left(\omega_{i}\right) E_{\beta}\left(\omega_{j}\right) e^{-i\left(\omega_{i}+\omega_{j}\right) t},
	\end{aligned}  \end{equation}
	where repeated indices $ \omega_{i}$, $\omega_{j} $, $ \alpha $ and $ \beta $ indicate summation \cite{Sipe2000}.
	We can obtain
	%	\begin{widetext}
	\begin{equation} \label{j-2}
		\begin{aligned}
			\left. \langle j_{\eta} \right. \rangle^{(2)}
			\equiv &\left[\sigma _{(2)\left( i \right) \left( ii \right)}^{\eta \alpha \beta}\left( \omega _i,\omega _j \right) +\sigma _{(2)\left( e \right) \left( ie \right)}^{\eta \alpha \beta}\left( \omega _i,\omega _j \right) \right.\\
			&+\sigma _{(2)\left( e \right) \left( ei \right)}^{\eta \alpha \beta}\left( \omega _i,\omega _j \right)+\sigma _{(2)\left( i \right) \left( ee \right)}^{\eta \alpha \beta}\left( \omega _i,\omega _j \right)\\
			& \left. +\sigma _{(2)\left( e \right) \left( ee \right)}^{\eta \alpha \beta}\left( \omega _i,\omega _j \right)\right]E_{\alpha}\left(\omega_{i}\right) E_{\beta}\left(\omega_{j}\right) e^{-i\left(\omega_{i}+\omega_{j}\right) t} ,\\
		\end{aligned}
	\end{equation}
	where
	\begin{equation} \begin{aligned}  \label{2nd-conductivity-i-ii}
			&\sigma _{(2)\left( i \right) \left( ii \right)}^{\eta \alpha \beta}\left( \omega _i,\omega _j \right)\\
			&=\frac{e^3}{\hbar ^3}\int{\frac{d^3k}{(2\pi )^3}} \sum_n{d_{}^{\omega _i}d_{}^{\omega _i+\omega _j}\left( \partial _{k_{\eta}}\varepsilon _n \right) \partial _{k_{\alpha}}\partial _{k_{\beta}}f_n(\vec{k})}   ,
	\end{aligned}  \end{equation}
	\begin{equation} \begin{aligned} \label{2nd-conductivity-e-ei}
			&\sigma _{(2)\left( e \right) \left( ei \right)}^{\eta \alpha \beta}\left( \omega _i,\omega _j \right)\\
			&=\frac{e^3}{\hbar ^3}\int{\frac{d^3k}{(2\pi )^3}} \sum_{nm}{d^{\omega _i}d_{nm}^{\omega _i+\omega _j}\varepsilon _{mn}\xi _{mn}^{\eta}\xi _{nm}^{\beta}\partial _{k_{\alpha}}f_{mn}(\vec{k})}  ,
	\end{aligned}  \end{equation}
	\begin{equation}   \label{2nd-conductivity-e-ie}
		\begin{aligned}
			&\sigma _{(2)\left( e \right) \left( ie \right)}^{\eta \alpha \beta}\left( \omega _i,\omega _j \right)\\
			&=\frac{e^3}{\hbar ^3}\int{\frac{d^3k}{(2\pi )^3}}\sum_{nm}{\left\{ d_{nm}^{\omega _i+\omega _j}\varepsilon _{mn}\xi _{mn}^{\eta}\xi _{nm}^{\alpha}\partial _{k_{\beta}}\left[ d_{nm}^{\omega _i}f_{mn}(\vec{k}) \right] \right.}\\
			&\quad\left.+d_{nm}^{\omega _i}d_{nm}^{\omega _i+\omega _j}\varepsilon _{mn}f_{mn}\xi _{mn}^{\eta}\xi _{nm;\beta}^{\alpha} \right\},\\
		\end{aligned}
	\end{equation}
	
	\begin{equation}  \label{2nd-conductivity-i-ee}   \begin{aligned}
			&\sigma _{(2)\left( i \right) \left( ee \right)}^{\eta \alpha \beta}\left( \omega _i,\omega _j \right)\\
			&=-\frac{e^3}{\hbar ^3}\int{\frac{d^3k}{(2\pi )^3}} \sum_{nm}{\hbar d_{}^{\omega _i+\omega _j}d_{nm}^{\omega _i}f_{nm}\Delta _{nm}^{\eta}\xi _{nm}^{\alpha}\xi _{mn}^{\beta}}  ,
	\end{aligned}  \end{equation}
	\begin{equation}   \label{2nd-conductivity-e-ee}
		\begin{aligned}
			&\sigma _{(2)\left( e \right) \left( ee \right)}^{\eta \alpha \beta}\left( \omega _i,\omega _j \right)\\
			&=-\frac{e^3}{\hbar ^3}\int{\frac{d^3k}{(2\pi )^3}}
			\sum_{nm}{\left[id_{nm}^{\omega _i+\omega _j}\varepsilon _{mn}\xi _{mn}^{\eta} \right.}\\
			&\quad\left.\times\sum_{l\ne n,m}{\left( d_{lm}^{\omega _i}\xi _{lm}^{\alpha}\xi _{nl}^{\beta}f_{ml}-d_{nl}^{\omega _i}\xi _{nl}^{\alpha}\xi _{lm}^{\beta}f_{ln} \right)} \right] ,
	\end{aligned}  \end{equation}
	%	\end{widetext}
	\begin{equation}
		d_{n m}^{\omega}\equiv \frac{1}{\omega-\varepsilon_{n m} / \hbar+i \Gamma},
	\end{equation}
	\begin{equation}
		d^{\omega}\equiv \frac{1}{\omega+i \Gamma},
	\end{equation}
	\begin{equation}\label{Delta_nm} \begin{aligned}
			\Delta_{n m}^{\alpha} \equiv v_{n n}^{\alpha}-v_{m m}^{\alpha}=\frac{1}{\hbar} \frac{\partial \varepsilon_{n m}}{\partial k_{\alpha}},
	\end{aligned}  \end{equation}
	and the ``generalized derivatives'' of $ \vec{\xi}_{ nm} $ is defined by \cite{Sipe2000}
	\begin{equation} \begin{aligned} \label{gen-der-xi}
			\xi_{n m ; \alpha}^{\beta}\equiv \partial_{k_{\alpha}} \xi_{n m}^{\beta}-i\left(\xi_{n n}^{\alpha}-\xi_{m m}^{\alpha}\right) \xi_{n m}^{\beta}.
	\end{aligned}  \end{equation}
	
	%\par The subscripts $ (i) $ and $ (ii) $ in Eq.~(\ref{2nd-conductivity-i-ii}) indicate that this term comes from the product of $ {v}_{  m n}^{(i)\eta} $ and $ \rho _{nm}^{(2)(ii)} $, and so on for other terms.
	\textcolor{red}{The subscripts ``$ i $'' and ``$ e $'' in Eq.~(\ref{j-2}) denote intraband and interband, respectively. Our classification of intraband and interband effects is based on how many times the intraband position operator $ \hat{\vec{r}}_i $ and interband position operator $ \hat{\vec{r}}_e $ are used in the derivation \cite{Aversa1995,Watanabe2021,Sipe2000,supplemental_material}. ``$ (i)(ii) $'' in Eq.~(\ref{2nd-conductivity-i-ii}) indicates that this term comes from the product of intraband velocity operator matrix element $ {v}_{  m n}^{(i)\eta} $ and intraband second-order density matrix $ \rho _{nm}^{(2)(ii)} $. ``$ (e)(ie) $'' in Eq.~(\ref{2nd-conductivity-e-ie}) indicates that this term comes from the product of interband velocity operator matrix element $ {v}_{  m n}^{(e)\eta} $ and intraband-interband mixing second-order density matrix $ \rho _{nm}^{(2)(ie)} $, and so on for other terms. }
	Hence $ \sigma _{(2)\left( i \right) \left( ii \right)}^{\eta \alpha \beta} $ is a pure intraband term and is also called the Drude term \cite{Watanabe2021,Lihm2022}. $ \sigma _{(2)\left( e \right) \left( ee \right)}^{\eta \alpha \beta} $ is a pure interband term, while the other second-order conductivity terms mix intraband and interband effects. Eqs.~(\ref{2nd-conductivity-i-ii})-(\ref{2nd-conductivity-e-ee}) correspond to Eqs.~(25)-(28) of Ref.~\onlinecite{Watanabe2021}, which have set $ \Gamma=0^{+} $ but here we do not and we give a more complete analysis of the intraband and interband effects.
	%, but we do not set $ \Gamma=0^{+} $.
	\par Now we discuss $ \sigma _{(2)(e)(ei)}^{\eta \alpha \beta} $. When $ \omega_{i}+\omega_{j} \ll |\varepsilon_{nm}/ \hbar | $ \cite{Matsyshyn2019} and $ \Gamma \ll |\varepsilon_{nm}/ \hbar | $, making use of
	\begin{equation}
		\frac{\varepsilon_{m n}}{\omega_{i}+\omega_{j}-\varepsilon_{n m} / \hbar+i \Gamma} \approx \frac{\varepsilon_{m n}}{\varepsilon_{m n} / \hbar+i \Gamma} \approx \hbar,
	\end{equation}
	and the definition of Berry curvature
	\begin{equation}\label{definition of Berry curvature}
		\begin{aligned}
			\Omega_{n}^{\alpha \eta} & \equiv \frac{\partial}{\partial k_{\alpha}} \xi_{n n}^{\eta}-\frac{\partial}{\partial k_{\eta}} \xi_{n n}^{\alpha},\\
			\Omega_n^z& \equiv\frac{\partial}{\partial {k_x}} \xi_{n n}^y-\frac{\partial}{\partial {k_y}} \xi_{n n}^x=\Omega_n^{x y}=-\Omega_n^{y x},
		\end{aligned}
	\end{equation}
	Eq.~(\ref{2nd-conductivity-e-ei}) can be reduced to the form of the Berry curvature dipole \cite{Matsyshyn2019}:
	\begin{equation}\label{traditional_BCD_formula} \begin{aligned}
			\sigma _{(2)(e)(ei)}^{\eta \alpha \beta}\left( \omega _i,\omega _j \right) =\frac{e^3}{\hbar ^2}\frac{i}{\omega _i+i\Gamma}\int{\frac{d^3k}{(2\pi )^3}}\sum_n{f_n(\vec{k})}\partial _{k_{\alpha}}\Omega _{n}^{\eta\beta },	
	\end{aligned}  \end{equation}
	so $ \sigma _{(2)\left( e \right) \left( ei \right)}^{\eta \alpha \beta}\left( \omega _i,\omega _j \right) $ is also called the BCD or ``interband 1'' term \cite{Matsyshyn2019}. For PGE, if $ \Gamma \ll |\varepsilon_{nm}/ \hbar | $ holds, Eq.~(\ref{2nd-conductivity-e-ei}) can reduce to the traditional BCD formula Eq.~(\ref{traditional_BCD_formula}) for light of any frequency. For SHG, Eq.~(\ref{traditional_BCD_formula}) holds only if both the low frequencies \textcolor{red}{($ 2\omega \ll |\varepsilon_{nm}/ \hbar | $)} and $ \Gamma \ll |\varepsilon_{nm}/ \hbar | $ are satisfied. However, when the relaxation time is taken as 5 ps \cite{Zheng2016} and 10 fs \cite{Aoki2022,Ma2018,Xu2018,Qin2021}, $\hbar \Gamma$ is 0.13 meV and 65.82 meV, respectively, which do not satisfy $ \Gamma \ll |\varepsilon_{nm}/ \hbar | $ for the $ T_d$-WTe$_2$ monolayer. For example, when the out-of-plane electric field is 0.2 V/nm, the maximum splitting value between the two lowest conduction bands near the gap-opening points is 26.1 meV \cite{Xu2018}. Thus we still use Eq.~(\ref{2nd-conductivity-BCD}) instead of Eq.~(\ref{traditional_BCD_formula}) to do the numerical calculation.
	\par Next we discuss $ \sigma _{(2)\left( i \right) \left( ee \right)}^{\eta \alpha \beta} $. %Making use of Eqs.~(\ref{Delta_nm}), one can easily prove that (see SM \cite{supplemental_material})
	%	\begin{equation}  \label{2nd-conductivity-i-ee-another1}
	%		\begin{aligned}
	%			&\sigma _{(2)\left( i \right) \left( ee \right)}^{\eta \alpha \beta}\left( \omega _i,\omega _j \right) \\
	%			&\quad =-\frac{e^3}{\hbar ^3}\int{\frac{d^3k}{(2\pi )^3}} \sum_{nm}{\hbar d_{}^{\omega _i+\omega _j}d_{nm}^{\omega _i}f_{nm}\Delta _{nm}^{\eta}\xi _{nm}^{\alpha}\xi _{mn}^{\beta}}  ,\\
	%		\end{aligned}
	%	\end{equation}
	%    where
	%\begin{equation}\label{Delta_nm} \begin{aligned}
	%		\Delta_{n m}^{\alpha} \equiv v_{n n}^{\alpha}-v_{m m}^{\alpha}=\frac{1}{\hbar} \frac{\partial \varepsilon_{n m}}{\partial k_{\alpha}} .
	%\end{aligned}  \end{equation}
	From Eq.~(\ref{j-2-define}) we know that \cite{Sipe2000}
	% and repeated indices indicate summation
	\begin{equation}\label{repeated-sigma-1}
		\begin{aligned}
			\left. \langle j_{\eta} \right. \rangle^{(2)} =&\sum_{\alpha \beta} \sum_{\omega_i \omega_j} \sigma_{(2)}^{\eta \alpha \beta}\left(\omega_i, \omega_j\right) E^\alpha\left(\omega_i\right) E^\beta\left(\omega_j\right) e^{-i\left(\omega_i+\omega_j\right) t} \\
			=&\sum_{\beta \alpha} \sum_{\omega_j \omega_i} \sigma_{(2)}^{\eta \beta\alpha}\left(\omega_j, \omega_i\right) E^\beta\left(\omega_j\right) E^\alpha\left(\omega_i\right) e^{-i\left(\omega_i+\omega_j\right) t} \\
			=&\sum_{\alpha \beta} \sum_{\omega_i \omega_j}\left[\frac{1}{2} \sigma_{(2)}^{\eta \alpha \beta}\left(\omega_i, \omega_j\right)+\frac{1}{2} \sigma_{(2)}^{\eta \beta \alpha}\left(\omega_j, \omega_i\right)\right]\\
			&\times E^\alpha\left(\omega_i\right) E^\beta\left(\omega_j\right) e^{-i\left(\omega_i+\omega_j\right) t}.
		\end{aligned}
	\end{equation}
	Note that we cannot conclude from Eq.~(\ref{repeated-sigma-1}) that $ \sigma _{(2)}^{\eta \alpha \beta}\left( \omega _i,\omega _j \right)=\sigma _{(2)}^{\eta  \beta\alpha}\left( \omega _j,\omega _i \right) $ (see SM \cite{supplemental_material}). However, if we define a new symmetrized second-order conductivity
	\begin{equation}  \label{new-2nd-conductivity}\begin{aligned}
			\sigma_{(2)\text{new}}^{\eta \alpha \beta}\left(\omega_i, \omega_j\right)=\frac{1}{2} \sigma_{(2)}^{\eta \alpha \beta}\left(\omega_i, \omega_j\right)+\frac{1}{2} \sigma_{(2)}^{\eta \beta \alpha}\left(\omega_j, \omega_i\right),	
	\end{aligned}  \end{equation}
	then
	\begin{equation}\label{relation-new-2nd-conductivity} \begin{aligned}
			\sigma_{(2)\text{new}}^{\eta \alpha \beta}\left(\omega_i, \omega_j\right) =\sigma_{(2)\text{new}}^{\eta \beta\alpha}\left(\omega_j, \omega_i\right)	  .
	\end{aligned}  \end{equation}
	\par Thus we can use Eq.~(\ref{new-2nd-conductivity}) to obtain another expression of $ \sigma _{(2)\left( i \right) \left( ee \right)}^{\eta \alpha \beta}\left( \omega _i,\omega _j \right) $ \cite{Lihm2022} (see SM \cite{supplemental_material})
	%\begin{widetext}
	\begin{equation}    \label{2nd-conductivity-i-ee-another2}
		\begin{aligned}
			&\sigma _{(2)\left( i \right) \left( ee \right)}^{\eta \alpha \beta}\left( \omega _i,\omega _j \right)\\
			&=\frac{-e^3}{\hbar ^3}\int{\frac{d^3k}{(2\pi )^3}} \sum_{nm}{\hbar d_{}^{\omega _i+\omega _j}d_{nm}^{\omega _i}f_{nm}\Delta _{nm}^{\eta}\xi _{nm}^{\alpha}\xi _{mn}^{\beta}}\\
			&\quad+[(\alpha,\omega_{i})\leftrightarrow (\beta,\omega_{j})]\\
			&=\frac{-e^3}{2\hbar ^3}\int{\frac{d^3k}{8\pi ^3}} \sum_{nm}{\hbar d_{}^{\omega _i+\omega _j}\left( d_{nm}^{\omega _i}+d_{mn}^{\omega _j} \right) f_{nm}\Delta _{nm}^{\eta}\xi _{mn}^{\beta}\xi _{nm}^{\alpha}} .\\
	\end{aligned}\end{equation}
	%\end{widetext}
	In the case of PGE and the clean limit ($ \Gamma \rightarrow 0^{+} $), Eq.~(\ref{2nd-conductivity-i-ee-another2}) can be reduced to a formula containing the expression of the injection current in Ref.~\onlinecite{Sipe2000} \cite{Watanabe2021,Lihm2022}. Therefore Eq.~(\ref{2nd-conductivity-i-ee-another2}) can also be called the ``injection'' or ``interband 2 (IB2)'' term under the relaxation time approximation ($ \Gamma $ is finite) \cite{Matsyshyn2019}.
	\par Finally we discuss $ \sigma _{(2)\left( e \right) \left( ie \right)}^{\eta \alpha \beta} $ and $\sigma _{(2)\left( e \right) \left( ee \right)}^{\eta \alpha \beta} $. In the case of PGE and the clean limit ($ \Gamma \rightarrow 0^{+} $), $ \sigma _{(2)\left( e \right) \left( ie \right)}^{\eta \alpha \beta}\left( \omega _i,\omega _j \right) +\sigma _{(2)\left( e \right) \left( ee \right)}^{\eta \alpha \beta}\left( \omega _i,\omega _j \right)  $ can be reduced to a formula containing the expression of the shift current in Ref.~\onlinecite{Sipe2000} \cite{Watanabe2021,Lihm2022}. Therefore $  \sigma _{(2)\left( e \right) \left( ie \right)}^{\eta \alpha \beta}\left( \omega _i,\omega _j \right) +\sigma _{(2)\left( e \right) \left( ee \right)}^{\eta \alpha \beta}\left( \omega _i,\omega _j \right)   $ can also be called the ``shift'' or ``interband 3 (IB3)'' term under the relaxation time approximation \cite{Matsyshyn2019}.
	% by using Eqs.~(\ref{2nd-conductivity-e-ie}) and (\ref{2nd-conductivity-e-ee})
	\par In summary, the second-order conductivity under the relaxation time approximation is \cite{Matsyshyn2019}
	%in the case of considering scattering effects ($ \Gamma $ is finite)
	\begin{equation} \begin{aligned}    \label{2nd-conductivity-another}
			\sigma _{(2)}^{\eta \alpha \beta}\left( \omega _i,\omega _j \right)&=\sigma _{(2)\mathrm{Drude}}^{\eta \alpha \beta}\left( \omega _i,\omega _j \right) +\sigma _{(2)\mathrm{BCD}}^{\eta \alpha \beta}\left( \omega _i,\omega _j \right) \\
			&\quad+\sigma _{(2)\mathrm{IB2}}^{\eta \alpha \beta}\left( \omega _i,\omega _j \right) +\sigma _{(2)\mathrm{IB3}}^{\eta \alpha \beta}\left( \omega _i,\omega _j \right)  ,
	\end{aligned}  \end{equation}
	where
	\begin{widetext}
		\begin{equation} \begin{aligned}  \label{2nd-conductivity-Drude}
				\sigma _{(2)\mathrm{Drude}}^{\eta \alpha \beta}\left( \omega _i,\omega _j \right)
				=&\frac{e^3}{2\hbar ^3}\int{\frac{d^3k}{(2\pi )^3}} \sum_n{d_{}^{\omega _i}d_{}^{\omega _i+\omega _j}\left( \partial _{k_{\eta}}\varepsilon _n \right) \partial _{k_{\alpha}}\partial _{k_{\beta}}f_n(\vec{k})} +[(\alpha,\omega_{i})\leftrightarrow (\beta,\omega_{j})] ,
		\end{aligned}  \end{equation}
		\begin{equation}\label{2nd-conductivity-BCD} \begin{aligned}
				\sigma _{(2)\mathrm{BCD}}^{\eta \alpha \beta}\left( \omega _i,\omega _j \right)
				=&\frac{e^3}{2\hbar ^3}\int{\frac{d^3k}{(2\pi )^3}} \sum_{nm}{d^{\omega _i}d_{nm}^{\omega _i+\omega _j}\varepsilon _{mn}\xi _{mn}^{\eta}\xi _{nm}^{\beta}\partial _{k_{\alpha}}f_{mn}(\vec{k})} +[(\alpha,\omega_{i})\leftrightarrow (\beta,\omega_{j})] ,
		\end{aligned}  \end{equation}
		\begin{equation}    \label{2nd-conductivity-IB2-injection}
			\begin{aligned}
				\sigma _{(2)\mathrm{IB2}}^{\eta \alpha \beta}\left( \omega _i,\omega _j \right)
				=&-\frac{e^3}{4\hbar ^3}\int{\frac{d^3k}{8\pi ^3}} \sum_{nm}{\hbar d_{}^{\omega _i+\omega _j}\left( d_{nm}^{\omega _i}+d_{mn}^{\omega _j} \right) f_{nm}\Delta _{nm}^{\eta}\xi _{mn}^{\beta}\xi _{nm}^{\alpha}} +[(\alpha,\omega_{i})\leftrightarrow (\beta,\omega_{j})] ,
		\end{aligned}\end{equation}
		\begin{equation}    \label{2nd-conductivity-IB3-shift}
			\begin{aligned}
				\sigma_{(2) \mathrm{IB} 3}^{\eta \alpha \beta}\left(\omega_{i}, \omega_{j}\right)=& \frac{e^{3}}{2 \hbar^{3}} \int \frac{d^{3} k}{(2 \pi)^{3}} \sum_{n m}\left\{d_{n m}^{\omega_{i}+\omega_{j}} \varepsilon_{m n} \xi_{m n}^{\eta} \xi_{n m}^{\alpha} \partial_{k_{\beta}}\left[d_{n m}^{\omega_{i}} f_{m n}(\vec{k})\right]+d_{n m}^{\omega_{i}} d_{n m}^{\omega_{i}+\omega_{j}} \varepsilon_{m n} f_{m n} \xi_{m n}^{\eta} \xi_{n m ; \beta}^{\alpha}\right.\\
				&\left.-i d_{n m}^{\omega_{i}+\omega_{j}} \varepsilon_{m n} \xi_{m n}^{\eta} \sum_{l \neq n, m}\left(d_{l m}^{\omega_{i}} \xi_{l m}^{\alpha} \xi_{n l}^{\beta} f_{m l}-d_{n l}^{\omega_{i}} \xi_{n l}^{\alpha} \xi_{l m}^{\beta} f_{l n}\right)\right\} +\left[\left(\alpha, \omega_{i}\right) \leftrightarrow\left(\beta, \omega_{j}\right)\right], %\\
				%			&+\left[\left(\alpha, \omega_{i}\right) \leftrightarrow\left(\beta, \omega_{j}\right)\right],
		\end{aligned}\end{equation}
	\end{widetext}
	which are all new second-order conductivities obtained from Eq.~(\ref{new-2nd-conductivity}). And we do not explicitly write the subscript ``new" for brevity. It can be proved that when the system has time-reversal symmetry, only the Drude term vanishes and all other terms still exist (see SM \cite{supplemental_material}). \textcolor{red}{In the Section V of SM \cite{supplemental_material}, using the phenomenological expressions for PGE in Section \ref{phenomenological expressions of photogalvanic effect}, we obtain the universal frequency dependence of the BCD term for the LPGE and CPGE cases as $ \operatorname{Re}(\sigma_{(2) \mathrm{BCD}}^{\eta \alpha \beta}(\omega,-\omega)) \propto 1 /(\omega^2+\Gamma^2) $ and $\operatorname{Im}(\sigma_{(2) \mathrm{BCD}}^{\eta \alpha \beta}(\omega,-\omega)) \propto \omega /(\omega^2+\Gamma^2) $, respectively. In the case of PGE and the system with time-reversal symmetry, we combine the phenomenological expressions to obtain the reduced expressions for the BCD and IB2 terms, and also find that the LPGE current induced by the BCD term exists only in the direction transverse to the alternating electric field; we also point out that the IB2 term exists only under circularly polarized light and introduce the physical meaning of IB2 term, i.e., when $ \Gamma $ is not reasonably small, the ``injection'' phenomenon will disappear due to the scattering effect and the injection current will saturate to the IB2 term \cite{Watanabe2021,Dai2023,Juan2017}.} However, the IB3 term is too complicated and it needs further study in the future. Note that in the case of 2D materials, ${d^3 k}/{ (2 \pi)^3}$ should be changed to ${d^2 k}/{(2 \pi)^2}$.
	
	%	We obtain the universal frequency dependence of the BCD term for the LPGE and CPGE cases as $ \operatorname{Re}\left(\sigma_{(2) \mathrm{BCD}}^{\eta \alpha \beta}(\omega,-\omega)\right) \propto 1 /\left(\omega^2+\Gamma^2\right) $ and $\operatorname{Im}\left(\sigma_{(2) \mathrm{BCD}}^{\eta \alpha \beta}(\omega,-\omega)\right) \propto \omega /\left(\omega^2+\Gamma^2\right) $, respectively.
	%the current induced by the BCD term in the LPGE case
	%In the case of PGE, with the use of the time-reversal symmetry and the phenomenological expressions of PGE in Section \ref{phenomenological expressions of photogalvanic effect}, the reduced expressions for the BCD and IB2 terms are given in the Section V of SM \cite{supplemental_material}, and the physical meaning of the IB2 term is also explained, i.e., when $ \Gamma $ is not reasonably small, the ``injection'' phenomenon will disappear due to the scattering effect and the injection current will saturate to the IB2 term \cite{Watanabe2021,Dai2023,Juan2017}. However, the IB3 term is too complicated and it needs further study in the future. Note that in the case of 2D materials, ${d^3 k}/{ (2 \pi)^3}$ should be changed to ${d^2 k}/{(2 \pi)^2}$.

	There is a lot of literature discussing the second-order nonlinear optical responses in the clean limit case \cite{Aversa1995,Sipe2000,Watanabe2021,Lihm2022}. But in practical experiments, the effect of scattering on nonlinear optical responses is important \cite{Sipe2000,Holder2020,Xu2021,belinicher1982kinetic}, and the relaxation time measured in many optical and transport experiments are on the order of $ \text{picoseconds} $ or $ \text{femtoseconds} $ \cite{Zheng2016,Wang2018,Aoki2022,Xu2021}.
	For a comparison with experiments, we have to take into account the relaxation time and go beyond the clean limit, because these results use $ \frac{\varepsilon _{mn}/\hbar}{\varepsilon _{mn}/\hbar +i0^+}\approx 1 $ and the Sokhotski-Plemelj Formula $\frac{1}{\omega-\varepsilon_{ nm}/\hbar+i 0^{+}}=P \frac{1}{ \omega-\varepsilon_{ nm}/\hbar}-i \pi \delta\left(\omega-\varepsilon_{ nm}/\hbar \right) $ \cite{Watanabe2021,Lihm2022,Sipe2000}\textcolor{red}{(this formula holds only when $\left|\omega-\varepsilon_{n m} / \hbar\right| \gg \Gamma$)}, which do not hold when $ \Gamma $ is on the order of $ 10^{11} $ Hz \cite{Zheng2016} or $ 10^{14} $ Hz \cite{Aoki2022,Ma2018,Xu2018,Qin2021}.
	In contrast, in this work, the advantage of second-order conductivity Eq.~(\ref{2nd-conductivity-another}) is that it can be used to handle a variety of situations such as arbitrary and finite relaxation rate $ \Gamma $ \textcolor{red}{(the premise is that the degree of disorder does not render the relaxation time approximation and band theory invalid)}, PGE and SHG.

	For insulators, $ \partial _{\vec{k}}f_n(\vec{k}) $ vanishes because the Fermi energy is in the band gap and the Fermi surface disappears. Therefore, Drude and the BCD terms do not exist in insulators \cite{Aversa1995,Lihm2022}. In addition, the existence of $ \partial _{\vec{k}}f_n(\vec{k}) $ also means that the Drude and BCD terms are inherent in the effect of Fermi surface.

	\section{phenomenological expression of Photogalvanic effect}\label{phenomenological expressions of photogalvanic effect}
	
	\textcolor{red}{In this section, we improve the existing theory \cite{Sturman1992,Ganichev2006} of phenomenological expressions for PGE [see the discussion below Eq.~(\ref{j-PGE-phenomenological})].} Consider a monochromatic polarized light incident along the normal to a 2D material interface with $ z=0 $ in the $ xy $ plane. Any kind of polarized light propagating along the $ z $ direction can be expressed as the superposition of two linearly polarized lights with electric field vectors along the $ x $-axis and $ y $-axis, respectively \cite{Liang2018}. Thus the electric field at $ z=0 $ is \cite{Liang2018}
	\begin{equation}\label{monochromatic polarized light}
		\begin{aligned}
			\vec{{E}}(t)
			&=a_{x} \cos \left(\phi_{x}-\omega t\right) \vec{e}_{x}+a_{y} \cos \left(\phi_{y}-\omega t\right) \vec{e}_{y}\\
			&\equiv \vec{E}(\omega) e^{-i \omega t}+\vec{E}(-\omega) e^{i \omega t} ,\\
		\end{aligned}
	\end{equation}
	where the amplitudes $a_{x}, a_{y}$ are real numbers, $\phi_{x}, \phi_{y}$ are the initial phases of the $ x $-direction and the $ y $-direction electric fields at $ z=0 $, respectively. $ \vec{E}(\omega) $ and $ \vec{E}(-\omega) $ are the complex amplitudes of the electric field, whose expressions can be seen in SM \cite{supplemental_material}, and they satisfy \cite{Watanabe2021}
	%The complex amplitude of the electric field is defined as
	%	\begin{equation}\label{complex amplitude-define}
	%			\begin{aligned}
	%				&\vec{E}(\omega )\equiv \frac{1}{2}a_xe^{i\phi _x}\vec{e}_x+\frac{1}{2}a_ye^{i\phi _y}\vec{e}_y,\\
	%				&\vec{E}(-\omega )\equiv \frac{1}{2}a_xe^{-i\phi _x}\vec{e}_x+\frac{1}{2}a_ye^{-i\phi _y}\vec{e}_y,\\
	%			\end{aligned}
	%	\end{equation}
	%	One has
	%	\begin{equation}
	%		\begin{aligned}
	%			\vec{{E}}(t)
	%            \equiv \vec{E}(\omega) e^{-i \omega t}+\vec{E}(-\omega) e^{i \omega t},
	%		\end{aligned}
	%	\end{equation}
	%	where the amplitude $ E_0 $ of the electric field $\vec{{E}}(t) $ and the complex unit polarization vector \cite{Ganichev2006} $ \vec{\epsilon} $ are defined as
	%	\begin{equation}\label{epsilon-define}
	%		\begin{aligned}
	%			&E_0=\sqrt{a_{x}^{2}+a_{y}^{2}}, \\
	%			&\vec{\epsilon}=\frac{a_{x}e^{i \phi_x} }{\sqrt{a_{x}^{2}+a_{y}^{2}}} \vec{e}_{x}+\frac{a_{y} e^{i \phi_y}}{\sqrt{a_{x}^{2}+a_{y}^{2}}} \vec{e}_{y}.
	%		\end{aligned}
	%	\end{equation}
	%	The complex amplitude of the electric field is defined as
	%	\begin{equation}\label{complex amplitude-define}
	%		\begin{aligned}
	%			&\vec{E}(\omega) \equiv \frac{1}{2} E_{0} \vec{\epsilon}, \\
	%			&\vec{E}(-\omega) \equiv \frac{1}{2} E_{0} \vec{\epsilon}^{*}.
	%		\end{aligned}
	%	\end{equation}
	%Because the time-domain electric field is a real number, we have\cite{Watanabe2021}
	% the following conclusion about complex amplitudes
	%	which satisfy \cite{Watanabe2021}
	\begin{equation} \label{conclusion-complex-E}  \begin{aligned}
			&  \vec{E}(\omega)=\vec{E}^{*}(-\omega).
	\end{aligned} \end{equation}
	We set $\delta\equiv\phi_y-\phi_x$ as the initial phase difference. $ E_0=\sqrt{a_{x}^{2}+a_{y}^{2}} $ is the amplitude of the electric field $\vec{{E}}(t) $.
	In the following discussion, the precondition is the new second-order conductivity Eq.~(\ref{new-2nd-conductivity}), otherwise we would not be able to derive the following phenomenological expression. For convenience, we will no longer explicitly write the subscript ``new'' in the following content.
	%	\par The first relation about the second-order conductivity of the PGE is Eq.~(\ref{relation-new-2nd-conductivity}).
	\par The photocurrents can be described phenomenologically as an expansion in powers of the incident electric field \cite{Sturman1992,Ganichev2006}, so the lowest order DC current density can be written as
	\begin{equation} \label{j-dc-1} \begin{aligned}
			\left\langle j_{\eta}\right\rangle_{0 \omega}^{(2)}=
			&\sum_{\alpha \beta}{\sigma _{(2)}^{\eta \alpha \beta}}(\omega ,-\omega )E_{\alpha}(\omega )E_{\beta}(-\omega )\\
			&+\sum_{\alpha \beta}{\sigma _{(2)}^{\eta \alpha \beta}}(-\omega ,\omega )E_{\alpha}(-\omega )E_{\beta}(\omega ),\\
	\end{aligned}  \end{equation}
	which can also be known from Eq.~(\ref{j-2-define}).
	Because $ \left\langle j_{\eta}\right\rangle_{0 \omega}^{(2)} $ is a real number \cite{Sturman1992}, one has (see SM \cite{supplemental_material}):
	\begin{equation} \label{PGE-conclusion2}
		\sigma_{(2)}^{\eta \alpha \beta}(-\omega, \omega)=\left[\sigma_{(2)}^{\eta \alpha \beta}(\omega,-\omega)\right]^{*}, \\
	\end{equation}
	where $\alpha=x,y$ and $\beta=x,y$.
	Eqs.~(\ref{relation-new-2nd-conductivity}) and (\ref{PGE-conclusion2}) can be combined to give the relation:
	\begin{equation}\label{PGE-conclusion3} \begin{aligned}
			\sigma_{(2)}^{\eta \alpha \beta}(-\omega, \omega)&=\left[\sigma_{(2)}^{\eta \beta \alpha}(-\omega, \omega)\right]^{*},\\
			\sigma_{(2)}^{\eta \alpha \beta}(\omega,-\omega)&=\left[\sigma_{(2)}^{\eta \beta \alpha}(\omega,-\omega)\right]^{*}.
	\end{aligned}  \end{equation}
	We define \cite{Ahn2020}
	\begin{equation}\label{sigma-real-imag}
		\sigma_{(2) }^{ \eta \alpha \beta}(\omega,-\omega) \equiv \sigma_{(2)\mathrm{Re}}^{ \eta \alpha \beta}(\omega,-\omega)+i \sigma_{(2)\mathrm{Im}}^{ \eta \alpha \beta}(\omega,-\omega),
	\end{equation}
	where
	\begin{equation}
		\begin{aligned}
			&\sigma_{(2)\mathrm{Re}} ^{\eta \alpha \beta}(\omega,-\omega) \equiv \operatorname{Re}\left[\sigma_{(2)} ^{\eta \alpha \beta}(\omega,-\omega)\right],\\
			& \sigma_{(2)\mathrm{Im}}^{\eta \alpha \beta}(\omega,-\omega) \equiv \operatorname{Im}\left[\sigma_{(2)}^{\eta \alpha \beta}(\omega,-\omega)\right].
		\end{aligned}
	\end{equation}
	Eqs.~(\ref{sigma-real-imag}) and (\ref{PGE-conclusion3}) can be combined to give the relation:
	\begin{equation}\label{PGE-conclusion4}
		\begin{aligned}
			&\sigma_{(2)\mathrm{Re}} ^{\eta \alpha \beta}(\omega,-\omega)=\sigma_{(2)\mathrm{Re}} ^{\eta  \beta\alpha}(\omega,-\omega),\\
			& \sigma_{(2)\mathrm{Im}}^{\eta \alpha \beta}(\omega,-\omega) =-\sigma_{(2)\mathrm{Im}}^{\eta  \beta\alpha}(\omega,-\omega),
		\end{aligned}
	\end{equation}
	i.e., the real and imaginary parts of $\sigma_{(2) }^{ \eta \alpha \beta}(\omega,-\omega) $ are symmetric and anti-symmetric for the exchange of $\alpha$ and $\beta$, respectively.
	%	Eqs.~(\ref{sigma-real-imag}) and (\ref{PGE-conclusion2}) can be combined to give the fifth relation:
	%	\begin{equation}
	%		\begin{aligned}
	%			&\sigma_{(2)\mathrm{Re}}^{\eta \alpha \beta}(\omega,-\omega)=\sigma_{(2)\mathrm{Re}}^{\eta \alpha \beta}(-\omega, \omega), \\
	%			&\sigma_{(2)\mathrm{Im}}^{\eta \alpha \beta}(\omega,-\omega)=-\sigma_{(2)\mathrm{Im}}^{\eta \alpha \beta}(-\omega, \omega).
	%		\end{aligned}
	%	\end{equation}
	
	%\subsection{Phenomenological expression of current density for photogalvanic effect}
	Any tensor that is antisymmetric about a pair of indices can be expressed as the product of a low-order tensor and the Levi-Civita totally antisymmetric pseudotensor $ \varepsilon_{\gamma \alpha \beta} $ \cite{Sturman1992,Ganichev2006}, hence
	\begin{equation} \label{sigma-C-Kappa-mid}
		\begin{aligned}
			\sigma_{(2)\mathrm{Im}}^{ \eta \alpha \beta}(\omega,-\omega) \equiv \kappa^{\eta \gamma}(\omega,-\omega) \varepsilon_{\gamma \alpha \beta},
	\end{aligned}  \end{equation}
	where $\kappa^{\eta \gamma}(\omega,-\omega)$ is a real second-order tensor, $ \gamma $ refers to Cartesian component. We are able to use Eq.~(\ref{sigma-C-Kappa-mid}) to obtain the expression of $ \kappa^{\eta \gamma}(\omega,-\omega) $:
	\begin{equation}\label{kappa}
		\begin{aligned}
			&\kappa^{\eta x}(\omega,-\omega)=\sigma_{(2)\mathrm{Im}}^{\eta y z}(\omega,-\omega), \\
			&\kappa^{\eta y}(\omega,-\omega)=\sigma_{(2)\mathrm{Im}}^{\eta z x}(\omega,-\omega), \\
			&\kappa^{\eta z}(\omega,-\omega)=\sigma_{(2)\mathrm{Im}}^{\eta x y}(\omega,-\omega),
		\end{aligned}
	\end{equation}
	where $\eta=x,y,z$.
	\par Making use of Eqs.~(\ref{relation-new-2nd-conductivity}), (\ref{conclusion-complex-E}), (\ref{j-dc-1}), (\ref{PGE-conclusion4}), (\ref{sigma-C-Kappa-mid}), we obtain the phenomenological expression of current density for PGE
	% and $ (\vec{E}(\omega)  \times \vec{E}(-\omega) )_{\gamma}=\varepsilon_{\gamma \alpha \beta} {E_{\alpha}}(\omega) {E_{\beta}}(-\omega)  $
	\begin{equation} \label{j-PGE-phenomenological}
		\begin{aligned}
			\left. \langle j_{\eta} \right. \rangle _{0\omega}^{(2)}
			=&2\sum_{\alpha \beta}{\sigma _{(2)\mathrm{Re}}^{\eta \alpha \beta}}(\omega ,-\omega )\operatorname{Re}(E_{\alpha}(\omega) E_{\beta}(-\omega)) \\
			&+2\sum_{\gamma}{\kappa ^{\eta \gamma}}(\omega ,-\omega )i(\vec{E}(\omega )\times \vec{E}(-\omega ))_{\gamma}.\\
	\end{aligned}\end{equation}
	
	The difference between Eq.~(\ref{j-PGE-phenomenological}) and the phenomenological expression of the PGE current in Refs.~\onlinecite{Sturman1992,Ganichev2006} \textcolor{red}{[see Eq. (1.9) of Ref.~\onlinecite{Sturman1992} and Eq. (7.5) of Ref.~\onlinecite{Ganichev2006}]} is the pre-factor 2 \textcolor{red}{in Eq.~(\ref{j-PGE-phenomenological})}. The reason for this difference is that the theory of Refs.~\onlinecite{Sturman1992,Ganichev2006} does not take into account that the repeated frequency index should also be summed when calculating the current density \textcolor{red}{[see Eq.~(\ref{j-2-define})], so the new second-order conductivity formula Eq.~(\ref{new-2nd-conductivity}) is also not considered.}
	
	%	Eqs.~(\ref{j-PGE-phenomenological}), (\ref{E-Pcirc-mid}), (\ref{P-circ}), (\ref{epsilon-define}) and (\ref{kappa}) lead to the phenomenological expression for the current density of the PGE under normal incidence:
	%	\begin{equation} \label{j-PGE-phenomenological-2}  \begin{aligned}
	%			\left. \langle j_{\eta} \right. \rangle _{0\omega}^{(2)}
	%			&=\frac{1}{2}\left\{ \sigma _{(2)\mathrm{Re}}^{\eta xx}(\omega ,-\omega )a_{x}^{2}+\sigma _{(2)\mathrm{Re}}^{\eta yy}(\omega ,-\omega )a_{y}^{2} \right.\\
	%			&+\left[\sigma _{(2)\mathrm{Re}}^{\eta xy}(\omega ,-\omega ) +\sigma _{(2)\mathrm{Re}}^{\eta yx}(\omega ,-\omega )\right]a_xa_y\cos \delta\\
	%			& \left. +\sigma _{(2)\mathrm{Im}}^{\eta xy}(\omega ,-\omega )P_{\text{circ}}E_{0}^{2} \right\}.\\
	%	\end{aligned} \end{equation}
	\par For linearly (L) polarized light, \textcolor{red}{the second line of Eq.~(\ref{j-PGE-phenomenological}) vanishes,} Eq.~(\ref{j-PGE-phenomenological}) reduces to
	\begin{equation} \begin{aligned} \label{linearly_polarized-PGE}
			& \left. \langle j_{\eta} \right. \rangle _{0\omega}^{(2)\left( \text{L} \right)} = \frac{1}{2}\left\{ \left. \sigma _{(2)\mathrm{Re}}^{\eta xx}(\omega ,-\omega )a_{x}^{2}+\sigma _{(2)\mathrm{Re}}^{\eta yy}(\omega ,-\omega )a_{y}^{2} \right.\right.\\
			&\left.+\left[ \sigma _{(2)\mathrm{Re}}^{\eta xy}(\omega ,-\omega )+\sigma _{(2)\mathrm{Re}}^{\eta yx}(\omega ,-\omega ) \right] a_xa_y\cos \delta \right\},  	
	\end{aligned}  \end{equation}
	where $\cos \delta=\pm1$. Only the real part of $ \sigma _{(2)}^{\eta \alpha \beta}(\omega ,-\omega ) $ contributes to $ \left. \langle j_{\eta} \right. \rangle _{0\omega}^{(2)\left( \text{L} \right)} $ \cite{Sturman1992,Ganichev2006}.
	\par For left-handed and right-handed circularly (LC and RC) polarized light, Eq.~(\ref{j-PGE-phenomenological}) reduces to
	\begin{equation}  \label{LC-PGE}  \begin{aligned}
			\left. \langle j_{\eta} \right. \rangle _{0\omega}^{(2)\left( \text{LC} \right)}=&\frac{1}{4}E_0^2\left[ \sigma _{(2)\mathrm{Re}}^{\eta xx}(\omega ,-\omega )\right.\\
			&\left.+\sigma _{(2)\mathrm{Re}}^{\eta yy}(\omega ,-\omega ) +2\sigma _{(2)\mathrm{Im}}^{\eta xy}(\omega ,-\omega ) \right],\\
	\end{aligned}\end{equation}
	\begin{equation} \label{RC-PGE}   \begin{aligned}
			\left. \langle j_{\eta} \right. \rangle _{0\omega}^{(2)\left( \text{RC} \right)}=&\frac{1}{4}E_0^2\left[ \sigma _{(2)\mathrm{Re}}^{\eta xx}(\omega ,-\omega )\right.\\
			&\left.+\sigma _{(2)\mathrm{Re}}^{\eta yy}(\omega ,-\omega )  -2\sigma _{(2)\mathrm{Im}}^{\eta xy}(\omega ,-\omega ) \right],\\
	\end{aligned}\end{equation}
	where $E_0^2$ represents the intensity of monochromatic polarized light \cite{Liang2018}. Although both $ \sigma _{(2)\mathrm{Re}}^{\eta \alpha \beta}(\omega ,-\omega ) $ and $ \sigma _{(2)\mathrm{Im}}^{\eta \alpha \beta}(\omega ,-\omega ) $ contribute to $ \left. \langle j_{\eta} \right. \rangle _{0\omega}^{(2)\left( \text{LC} \right)} $ and $ \left. \langle j_{\eta} \right. \rangle _{0\omega}^{(2)\left( \text{RC} \right)} $, the current difference between $ \left. \langle j_{\eta} \right. \rangle _{0\omega}^{(2)\left( \text{LC} \right)} $ and $ \left. \langle j_{\eta} \right. \rangle _{0\omega}^{(2)\left( \text{RC} \right)} $ \cite{Ahn2020} is
	\begin{equation}\label{CPGE_response} \begin{aligned}
			\left. \langle j_{\eta} \right. \rangle _{0\omega}^{(2)\left( \text{CPGE} \right)}&\equiv \left. \langle j_{\eta} \right. \rangle _{0\omega}^{(2)\left( \text{LC} \right)}-\left. \langle j_{\eta} \right. \rangle _{0\omega}^{(2)\left( \text{RC} \right)}\\
			&=\sigma _{(2)\mathrm{Im}}^{\eta xy}(\omega ,-\omega )E_{0}^{2},
	\end{aligned}  \end{equation}
	where we have defined $ 	\left. \langle j_{\eta} \right. \rangle _{0\omega}^{(2)\left( \text{CPGE} \right)} $ as the response of CPGE \cite{Xu2018}.
	Only the imaginary part of $ \sigma _{(2)}^{\eta xy}(\omega ,-\omega ) $ contributes to $ \left. \langle j_{\eta} \right. \rangle _{0\omega}^{(2)\left( \text{CPGE} \right)} $, \textcolor{red}{i.e., only the second line of Eq.~(\ref{j-PGE-phenomenological}) contributes to $ \left. \langle j_{\eta} \right. \rangle _{0\omega}^{(2)\left( \text{CPGE} \right)} $ \cite{Sturman1992,Ganichev2006,Glazov2014}.}
	%	\par For left-handed and right-handed elliptically (LE and RE) polarized light with ellipses of the same shape and size,
	%	\begin{equation} \begin{aligned}
	%			\left. \langle j_{\eta} \right. \rangle _{0\omega}^{(2)\left( \text{LE} \right)}-\left. \langle j_{\eta} \right. \rangle _{0\omega}^{(2)\left( \text{RE} \right)}=\sigma _{(2)\mathrm{Im}}^{\eta xy}(\omega ,-\omega )P_{\text{circ}}^{\left( \text{LE} \right)}E_{0}^{2}.
	%	\end{aligned}  \end{equation}

	The above content is phenomenological, however Eqs.~(\ref{2nd-conductivity-Drude})-(\ref{2nd-conductivity-IB3-shift}) are obtained by the quantum kinetics method, so we need to verify whether they satisfy \textcolor{red}{the phenomenological relation Eq.~(\ref{PGE-conclusion2})}, and in SM \cite{supplemental_material} we prove that they do. Therefore, we can directly take the real and imaginary parts of Eqs.~(\ref{2nd-conductivity-Drude})-(\ref{2nd-conductivity-IB3-shift}) in the case of PGE to obtain the LPGE and CPGE conductivities.
	%Note that the time-reversal symmetry is not used in the verification process.

	\section{phenomenological expressions of Second harmonic generation}
	%\subsection{Five relations for the second-order conductivity of the second harmonic generation}\label{Five conclusions-SHG}
	\textcolor{red}{In this section, we present a phenomenological analysis for LSHG and CSHG.} In the following discussion, the precondition is also the new second-order conductivity Eq.~(\ref{new-2nd-conductivity}).
	%The first relation about the second-order conductivity of the SHG is Eq.~(\ref{relation-new-2nd-conductivity}).
	Phenomenologically, the lowest order second harmonic current density can be written as
	\begin{equation} \label{j-SHG-1}  \begin{aligned}
			\left. \langle j_{\eta} \right. \rangle _{2\omega}^{(2)}=&\sum_{\alpha \beta}{\sigma _{(2)}^{\eta \alpha \beta}}(\omega ,\omega )E_{\alpha}(\omega )E_{\beta}(\omega )e^{-2i\omega t}\\
			&+\sum_{\alpha \beta}{\sigma _{(2)}^{\eta \alpha \beta}}(-\omega ,-\omega )E_{\alpha}(-\omega )E_{\beta}(-\omega )e^{2i\omega t},\\
	\end{aligned} \end{equation}
	which can also be known from Eq.~(\ref{j-2-define}).
	From the fact that $ \left\langle j_{\eta}\right\rangle_{2 \omega}^{(2)} $ is a real number \cite{Sturman1992} and Eq.~(\ref{conclusion-complex-E}), one has the relation (see SM \cite{supplemental_material}):
	\begin{equation}\label{SHG-conclusion2} \begin{aligned}
			\sigma _{(2)}^{\eta \alpha \beta}(\omega ,\omega )=\left[ \sigma _{(2)}^{\eta \alpha \beta}(-\omega ,-\omega ) \right] ^*,
	\end{aligned}  \end{equation}
	where $\alpha=x,y$ and $\beta=x,y$.
	Eqs.~(\ref{relation-new-2nd-conductivity}) and (\ref{SHG-conclusion2}) can be combined to give relation:
	\begin{equation} \begin{aligned}
			\sigma _{(2)}^{\eta \alpha \beta}(\omega ,\omega )=\left[ \sigma _{(2)}^{\eta \beta \alpha}(-\omega ,-\omega ) \right] ^*.
	\end{aligned}  \end{equation}
	We define
	\begin{equation}\label{SHG-sigma-real-imag}
		\sigma_{(2) }^{ \eta \alpha \beta}(\omega,\omega) \equiv \sigma_{(2) \mathrm{Re}}^{ \eta \alpha \beta}(\omega,\omega)+i \sigma_{(2) \mathrm{Im}}^{ \eta \alpha \beta}(\omega,\omega),
	\end{equation}
	where
	\begin{equation}
		\begin{aligned}
			&\sigma_{(2) \mathrm{Re}}^{ \eta \alpha \beta}(\omega,\omega)\equiv \operatorname{Re}\left[\sigma_{(2)} ^{\eta \alpha \beta}(\omega,\omega)\right],\\
			& \sigma_{(2) \mathrm{Im}}^{\eta \alpha \beta}(\omega,\omega) \equiv \operatorname{Im}\left[\sigma_{(2)}^{\eta \alpha \beta}(\omega,\omega)\right].
		\end{aligned}
	\end{equation}
	Eqs.~(\ref{SHG-sigma-real-imag}) and (\ref{relation-new-2nd-conductivity}) can be combined to give the relation:
	\begin{equation}\label{SHG-conclusion4}
		\begin{aligned}
			\sigma_{(2) \mathrm{Re}}^{\eta \alpha \beta}(\omega, \omega) &=\sigma_{(2) \mathrm{Re}}^{\eta \beta \alpha}(\omega,\omega) ,\\
			\sigma_{(2) \mathrm{Im}}^{\eta \alpha \beta}(\omega, \omega) &=\sigma_{(2) \mathrm{Im}}^{\eta \beta \alpha}(\omega,\omega),
		\end{aligned}
	\end{equation}
	i.e., the real and imaginary parts of $\sigma_{(2) }^{ \eta \alpha \beta}(\omega,\omega) $ are all symmetric for the exchange of $\alpha$ and $\beta$.
	%	Eqs.~(\ref{SHG-sigma-real-imag}) and (\ref{SHG-conclusion2}) can be combined to give the fifth relation:
	%	\begin{equation}\label{SHG-conclusion5}
	%		\begin{aligned}
	%			\sigma_{(2) \mathrm{Re}}^{\eta \alpha \beta}(\omega, \omega) &=\sigma_{(2) \mathrm{Re}}^{\eta\alpha \beta }(-\omega,-\omega) ,\\
	%			\sigma_{(2) \mathrm{Im}}^{\eta \alpha \beta}(\omega, \omega) &=-\sigma_{(2) \mathrm{Im}}^{\eta\alpha \beta }(-\omega,-\omega).
	%		\end{aligned}
	%	\end{equation}
	\par One can also easily verify that Eqs.~(\ref{2nd-conductivity-Drude})-(\ref{2nd-conductivity-IB3-shift}) satisfy the \textcolor{red}{phenomenological relation Eq.~(\ref{SHG-conclusion2})}.
	
	%\subsection{Phenomenological expression of current density for second harmonic generation}
	Eqs.~(\ref{j-SHG-1}) and (\ref{SHG-conclusion2}) lead to the phenomenological expression for the current density of the SHG under normal incidence (see SM \cite{supplemental_material}):
	%	\begin{equation}\label{j-SHG-3}
	%		\begin{aligned}
	%			\left\langle j_\eta\right\rangle_{2 \omega}^{(2)}
	%			=& 2 \operatorname{Re}\left(\sigma_{(2)}^{\eta x y}(\omega, \omega) E_x(\omega) E_y(\omega) e^{-2 i \omega t}\right) \\
	%			&+\operatorname{Re}\left(\sigma_{(2)}^{\eta x x}(\omega, \omega) E_x(\omega) E_x(\omega) e^{-2 i \omega t}\right)\\
	%			&+\operatorname{Re}\left(\sigma_{(2)}^{\eta y y}(\omega, \omega) E_y(\omega) E_y(\omega) e^{-2 i \omega t}\right).
	%		\end{aligned}
	%	\end{equation}
	%	Making use of Eqs. (\ref{epsilon-define}) and (\ref{complex amplitude-define}), Eq.~(\ref{j-SHG-3}) reduces to
	\begin{widetext}
		\begin{equation} \label{j-SHG-4}   \begin{aligned}
				\left. \langle j_{\eta} \right. \rangle _{2\omega}^{(2)}=&\frac{1}{2}a_xa_y\cos \left( \phi _x+\phi _y \right) \left[ \sigma _{(2)\mathrm{Re}}^{\eta xy}(\omega ,\omega )\cos\mathrm{(}2\omega t)+\sigma _{(2)\mathrm{Im}}^{\eta xy}(\omega ,\omega )\sin\mathrm{(}2\omega t) \right]\\
				&+\frac{1}{2}a_xa_y\sin \left( \phi _x+\phi _y \right) \left[ \sigma _{(2)\mathrm{Re}}^{\eta xy}(\omega ,\omega )\sin\mathrm{(}2\omega t)-\sigma _{(2)\mathrm{Im}}^{\eta xy}(\omega ,\omega )\cos\mathrm{(}2\omega t) \right]\\
				&+\frac{1}{4}a_{x}^{2}\cos \left( 2\phi _x \right) \left[ \sigma _{(2)\mathrm{Re}}^{\eta xx}(\omega ,\omega )\cos\mathrm{(}2\omega t)+\sigma _{(2)\mathrm{Im}}^{\eta xx}(\omega ,\omega )\sin\mathrm{(}2\omega t) \right]\\
				&+\frac{1}{4}a_{x}^{2}\sin \left( 2\phi _x \right) \left[ \sigma _{(2)\mathrm{Re}}^{\eta xx}(\omega ,\omega )\sin\mathrm{(}2\omega t)-\sigma _{(2)\mathrm{Im}}^{\eta xx}(\omega ,\omega )\cos\mathrm{(}2\omega t) \right]\\
				&+\frac{1}{4}a_{y}^{2}\cos \left( 2\phi _y \right) \left[ \sigma _{(2)\mathrm{Re}}^{\eta yy}(\omega ,\omega )\cos\mathrm{(}2\omega t)+\sigma _{(2)\mathrm{Im}}^{\eta yy}(\omega ,\omega )\sin\mathrm{(}2\omega t) \right]\\
				&+\frac{1}{4}a_{y}^{2}\sin \left( 2\phi _y \right) \left[ \sigma _{(2)\mathrm{Re}}^{\eta yy}(\omega ,\omega )\sin\mathrm{(}2\omega t)-\sigma _{(2)\mathrm{Im}}^{\eta yy}(\omega ,\omega )\cos\mathrm{(}2\omega t) \right].\\
		\end{aligned}\end{equation}
	\end{widetext}
	
	\begin{figure*}[t!]
		\begin{center}
			\includegraphics[width=2\columnwidth]{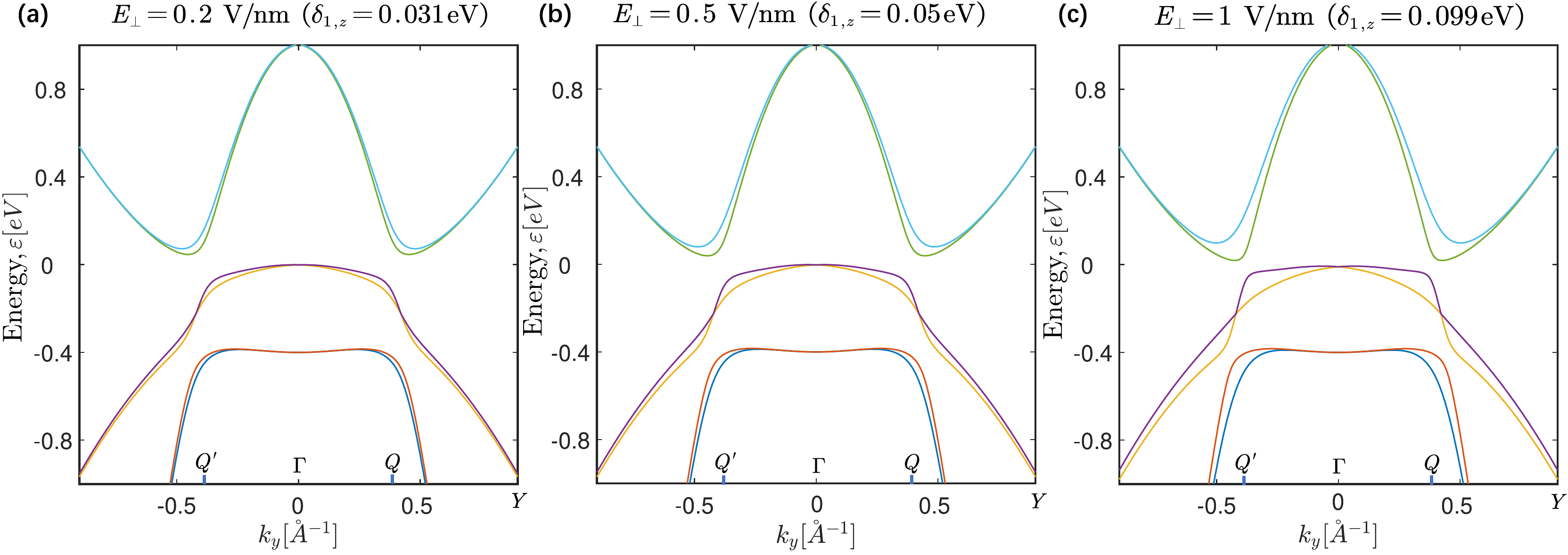}
		\end{center}
		\caption{ (a,b,c) Band structures of monolayer $ T_d$-WTe$_2$ with (a) $E_{\perp}=0.2 \mathrm{~V} / \mathrm{nm}$, (b) $E_{\perp}=0.5 \mathrm{~V} / \mathrm{nm}$ and (c) $E_{\perp}=1 \mathrm{~V} / \mathrm{nm}$ obtained by using the six-band model. $ Q $ and $ Q^{\prime}$ are the gap-opening points.} \label{Fig2}
	\end{figure*}  %Use the figure* environment to get a wide figure that spans the page in \texttt{twocolumn} formatting
	
	\par For linearly polarized light, if we assume that the alternating electric field is along the $ y $ direction, Eq.~(\ref{j-SHG-4}) reduces to
	\begin{equation}   \begin{aligned}
			\left. \langle j_{\eta} \right. \rangle _{2\omega}^{(2)\left( \text{L} \right)}
			=\sigma_{(2)\mathrm{eff}}^{\eta y y}(\omega, \omega) E_0^2 \sin \left(2 \omega t+\varphi_1\right),\\
	\end{aligned}\end{equation}
	where $E_0^2=a_y^2$ is the intensity of light, $ \tan \varphi _1\equiv\frac{\cos \left( 2\phi _x \right) \sigma _{(2)\mathrm{Re}}^{\eta yy}(\omega ,\omega )-\sin \left( 2\phi _x \right) \sigma _{(2)\mathrm{Im}}^{\eta yy}(\omega ,\omega )}{\cos \left( 2\phi _x \right) \sigma _{(2)\mathrm{Im}}^{\eta yy}(\omega ,\omega )+\sin \left( 2\phi _x \right) \sigma _{(2)\mathrm{Re}}^{\eta yy}(\omega ,\omega )} $ and we define the effective LSHG conductivity as
	\begin{equation}\label{LSHG-eff} \begin{aligned}
			\sigma_{(2)\mathrm{eff}}^{\eta y y}(\omega, \omega)\equiv \frac{1}{4}\sqrt{\left( \sigma _{(2)\mathrm{Re}}^{\eta yy}(\omega ,\omega ) \right) ^2+\left( \sigma _{(2)\mathrm{Im}}^{\eta yy}(\omega ,\omega ) \right) ^2}.
	\end{aligned}  \end{equation}
	We can see that the amplitude of the second harmonic current is independent of the initial phase $ \phi_x $.
	\par For left-handed and right-handed circularly polarized light, the current difference between $ \left. \langle j_{\eta} \right. \rangle _{2\omega}^{(2)\left( \text{LC} \right)} $ and $ \left. \langle j_{\eta} \right. \rangle _{2\omega}^{(2)\left( \text{RC} \right)} $ is
	\begin{equation}\label{CSHG-current} \begin{aligned}
			\left. \langle j_{\eta} \right. \rangle _{2\omega}^{(2)\left( \text{CSHG} \right)}&\equiv\left. \langle j_{\eta} \right. \rangle _{2\omega}^{(2)\left( \text{LC} \right)}-\left. \langle j_{\eta} \right. \rangle _{2\omega}^{(2)\left( \text{RC} \right)}\\
			& =\sigma_{(2)\mathrm{eff}}^{\eta x y}(\omega, \omega) E_0^2 \sin \left(2 \omega t+\varphi_2\right),
	\end{aligned}  \end{equation}
	where we have defined $ \left. \langle j_{\eta} \right. \rangle _{2\omega}^{(2)\left( \text{CSHG} \right)} $ as the response of CSHG, $\tan \varphi_2 \equiv\frac{\sin \left( 2\phi _x \right) \sigma _{(2)\mathrm{Re}}^{\eta xy}(\omega ,\omega )+\cos \left( 2\phi _x \right) \sigma _{(2)\mathrm{Im}}^{\eta xy}(\omega ,\omega )}{\sin \left( 2\phi _x \right) \sigma _{(2)\mathrm{Im}}^{\eta xy}(\omega ,\omega )-\cos \left( 2\phi _x \right) \sigma _{(2)\mathrm{Re}}^{\eta xy}(\omega ,\omega )}$ and we define the effective CSHG conductivity as
	\begin{equation}\label{CSHG-eff} \begin{aligned}
			\sigma_{(2)\mathrm{eff}}^{\eta x y}(\omega, \omega)\equiv \frac{1}{2}\sqrt{\left( \sigma _{(2)\mathrm{Re}}^{\eta xy}(\omega ,\omega ) \right) ^2+\left( \sigma _{(2)\mathrm{Im}}^{\eta xy}(\omega ,\omega ) \right) ^2}.
	\end{aligned}  \end{equation}
	The amplitude of $ \left. \langle j_{\eta} \right. \rangle _{2\omega}^{(2)\left( \text{CSHG} \right)} $ is independent of the initial phase $ \phi_x $.
	\par Unlike the PGE, for SHG, both $ \sigma _{(2)\mathrm{Re}}^{\eta \alpha \beta}(\omega ,\omega ) $ and $ \sigma _{(2)\mathrm{Im}}^{\eta \alpha \beta}(\omega ,\omega ) $ contribute to $ \left. \langle j_{\eta} \right. \rangle _{2\omega}^{(2)\left( \text{L} \right)} $ and $ \left. \langle j_{\eta} \right. \rangle _{2\omega}^{(2)\left( \text{CSHG} \right)} $.

	\section{Gatetunable second-order nonlinear optical responses from radio to infrared region in $ T_d-\mathrm{WTe}_2$ monolayer}
	
	\begin{figure*}[t!]
		\begin{center}
			\includegraphics[width=2\columnwidth]{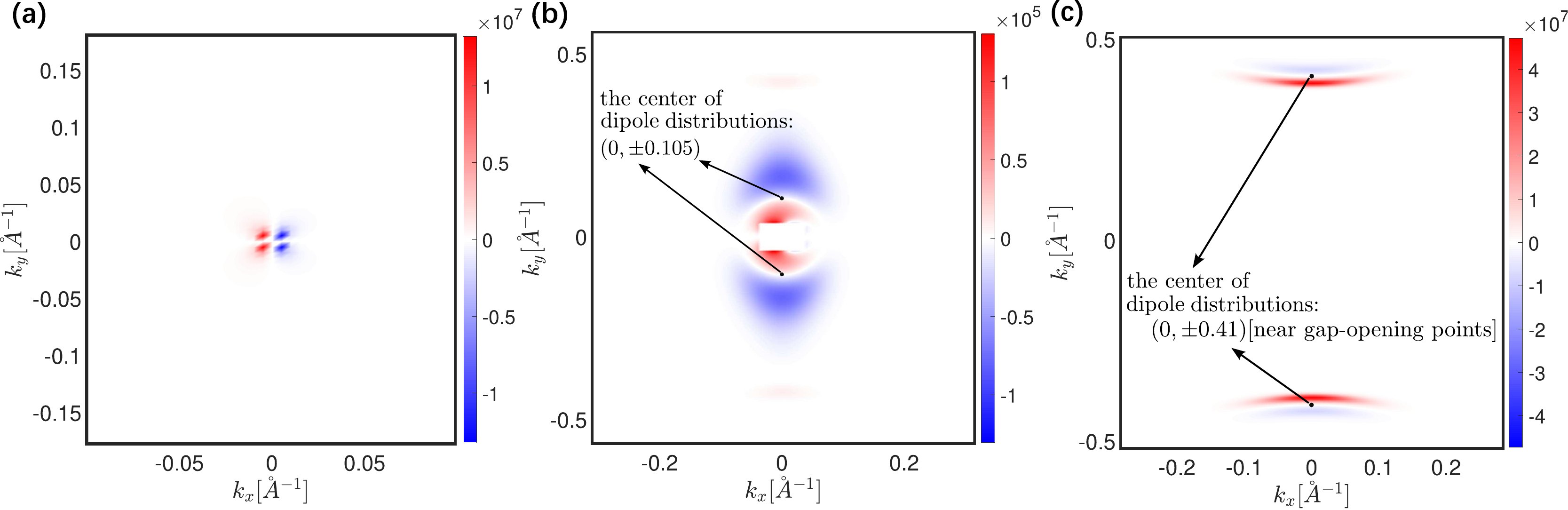}
		\end{center}
		\caption{ (Color online) (a,b,c) The $ \vec{k} $-space distribution of $I_{\mathrm{BCD}, \mathrm{LPGE}}^{x y y} $ for Fermi level (a,b) $\varepsilon _F=0 \mathrm{~eV}$ and (c) $\varepsilon _F=0.095 \mathrm{~eV}$ when the frequency $\nu = 1000 $ $ \text{Hz} $ and $E_{\perp}=0.2 \mathrm{~V} / \mathrm{nm}$. The difference between (a) and (b) is that the large magnitude of $ I_{\mathrm{BCD}, \mathrm{LPGE}}^{x y y} $ in the rectangular region near $ \Gamma $ point is discarded in (b). The unit of $I_{\mathrm{BCD}, \mathrm{LPGE}}^{x y y} $ is $ \text{\AA}^3/\text{eV} $.} \label{Fig3}
	\end{figure*}  %Use the figure* environment to get a wide figure that spans the page in \texttt{twocolumn} formatting
	The six-band model of the $ T_d$-WTe$_2$ monolayer in Ref.~\onlinecite{Shi2019} can capture the full reciprocal space distribution of geometric quantities such as Berry curvature, while the four-band model of Refs.~\onlinecite{Shi2019,Xu2018} can only describe the region near the gap opening points $ Q $ and $ Q^{\prime} $. The ab initio calculations show that the Berry curvature of the highest valence band actually has a large value at a point far from the gap opening points \cite{Xu2018}, which is a conclusion that the four-band model cannot capture but the six-band model can. In addition, when the frequency of light is appropriately high, more bands will be involved in the interband transitions \cite{Xu2018}, so the six-band model is more accurate than the four-band model. Therefore, considering the above reasons, we use the six-band model instead of the four-band model for the calculation.
	\par The six-band $ \vec{k}\cdot \vec{p} $ Hamiltonian of monolayer $ T_d$-WTe$_2$ with a bandgap of 55 meV is \cite{Shi2019}
	\begin{equation}\label{H_0-WTe2}
		H_{0}(\vec{k})=\left(\begin{array}{cccccc}
			\epsilon_{1} & v_{1}^{+} & 0 & 0 & 0 & 0 \\
			-v_{1}^{-} & \epsilon_{2} & v_{3}^{+} & 0 & 0 & 0 \\
			0 & -v_{3}^{-} & \epsilon_{3} & 0 & 0 & 0 \\
			0 & 0 & 0 & \epsilon_{1} & v_{1}^{-} & 0 \\
			0 & 0 & 0 & -v_{1}^{+} & \epsilon_{2} & v_{3}^{-} \\
			0 & 0 & 0 & 0 & -v_{3}^{+} & \epsilon_{3}
		\end{array}\right),
	\end{equation}
	where $\epsilon_i=c_{i, 0}+c_{i, x} k_x^2+c_{i, y} k_y^2$ and $v_i^{\pm}=\pm v_{i, x} k_x+i v_{i, y} k_y$ for the $ i $-th orbital. The values of parameters in Eq.~(\ref{H_0-WTe2}) are listed in Table III of Ref.~\onlinecite{Shi2019}. The applied out-of-plane electric field $E_{\perp}$ causes the total Hamiltonian to become
	\begin{equation}\label{H-WTe2} \begin{aligned}
			H(\vec{k})=H_{0}(\vec{k})+H_{1}(\vec{k}),
	\end{aligned}  \end{equation}
	where \cite{Shi2019}
	\begin{equation}
		H_{1}(\vec{k})=\left(\begin{array}{cccccc}
			0 & i \delta_{1, z} & 0 & 0 & i \delta_{1 ,x} & 0 \\
			-i \delta_{1, z} & 0 & i \delta_{3, z} & -i \delta_{1,x} & 0 & i \delta_{3 ,x} \\
			0 & -i \delta_{3, z} & 0 & 0 & -i \delta_{3, x} & 0 \\
			0 & i \delta_{1, x} & 0 & 0 & -i \delta_{1, z} & 0 \\
			-i \delta_{1, x} & 0 & i \delta_{3, x} & i \delta_{1, z} & 0 & -i \delta_{3, z} \\
			0 & -i \delta_{3, x} & 0 & 0 & i \delta_{3, z} & 0
		\end{array}\right),
	\end{equation}
	where $ \delta_{i, x} $ and $ \delta_{i, z} $ are connected to the electric field induced Rashba and Ising spin-orbit couplings, respectively, and they are both $ k $-independent. We approximately take $ \delta_{3, x} =\delta_{3, z} =0 $ because they originate from the third energy band far from the Fermi level \cite{Shi2019}. In addition, since $ \delta_{1, z} $ overwhelms $ \delta_{1, x} $, one can mainly use $ \delta_{1, z} $ to describe the spin-orbit coupling induced by $E_{\perp}$ \cite{Shi2019}. With the increase of $ \delta_{1, z} $, the spin splitting of the energy band increases gradually. By comparing with the splitting magnitude of the conduction bands near the gap-opening point at different electric fields calculated by ab initio in Ref.~\onlinecite{Xu2018}, we can know that the electric fields $E_{\perp}$ of $ 0.2 $ $\text{V/nm}$, $ 0.5 $ $\text{V/nm}$ and $ 1 $ $\text{V/nm}$ correspond to $(\delta_{1,x}, \delta_{1,z})$ = (0.01, 0.031) $ \text{eV}$, (0.01, 0.05) $ \text{eV}$ and (0.01, 0.099) $ \text{eV}$, respectively.
	\par We take the relaxation time $ \tau $ as 5 $ \text{ps} $, which is obtained from optical measurements \cite{Zheng2016}. The temperature is taken as $ 80 $ $ \text{K} $.
	
	\par The mirror symmetry $ \mathcal{M}_y $ of $ T_d$-WTe$_2$ monolayer leads to the vanishing of the components with an odd number of $ y $ indices in the second-order conductivity tensor \cite{Xu2018,Shi2019,Bhalla2022}, which is consistent with our following numerical calculations. Thus, from Eqs.~(\ref{LC-PGE}) and (\ref{RC-PGE}), for $ T_d$-WTe$_2$ monolayer, the currents of CPGE in $y$ and $x$ direction read
	\begin{equation}\label{LC-RC-L-0w-wte2} \begin{aligned}
			\left\langle j_y\right\rangle_{0 \omega}^{(2)(\mathrm{LC})}&=-\left\langle j_y\right\rangle_{0 \omega}^{(2)(\mathrm{RC})}=\frac{1}{2} \sigma_{(2) \mathrm{Im}}^{y x y}(\omega,-\omega) E_0^2, \\
			%		\left\langle j_y\right\rangle_{0 \omega}^{(2)(\mathrm{RC})}&=-\frac{1}{2} \sigma_{(2) \mathrm{Im}}^{y x y}(\omega,-\omega) E_0^2, \\
			\left\langle j_x\right\rangle_{0 \omega}^{(2)(\mathrm{LC})}&=\left\langle j_x\right\rangle_{0 \omega}^{(2)(\mathrm{RC})}\\
			&=\frac{1}{4}\left[\sigma_{(2) \mathrm{Re}}^{x x x}(\omega,-\omega)+\sigma_{(2) \mathrm{Re}}^{x y y}(\omega,-\omega)\right] E_0^2.
	\end{aligned}  \end{equation}
	Consequently, when the circularly polarized light changes from LC to RC, the DC current in the $ y $-direction will be reversed while the DC current in the $ x $-direction remains in the same direction, which is consistent with the experimental observations in Ref.~\onlinecite{Xu2018}. From Eq.~(\ref{j-SHG-4}), for $ T_d$-WTe$_2$ monolayer, the currents of CSHG in $y$ direction read
	\begin{equation}\label{LC-RC-2w-wte2}\begin{aligned}
			\left. \langle j_{y} \right. \rangle _{2\omega}^{(2)\left( \text{LC} \right)}&=-\langle j_{y} \rangle _{2\omega}^{(2)\left( \text{RC} \right)} \\ &=\frac{1}{2}\sigma_{(2)\mathrm{eff}}^{y x y}(\omega, \omega) E_0^2 \sin \left(2 \omega t+\varphi_2\right).
			%\\
			%		\left. \langle j_{y} \right. \rangle _{2\omega}^{(2)\left( \text{RC} \right)}&=-\frac{1}{2}\sigma_{(2)\mathrm{eff}}^{y x y}(\omega, \omega) E_0^2 \sin \left(2 \omega t+\varphi_2\right). \\
	\end{aligned}  \end{equation}

	\subsection{LPGE (or DC current of NHE) at 1000 $ \text{Hz} $}
	In experiments, Fermi level $\varepsilon _F$ and the out-of-plane electric field $E_{\perp}$ caused by the gate voltage can be controlled independently and they have significant influence on nonlinear optical responses and NHE \cite{Ma2018,Xu2018,Xiao2020}.

\begin{figure}[t!]
		\begin{center}
			\includegraphics[width=1\columnwidth]{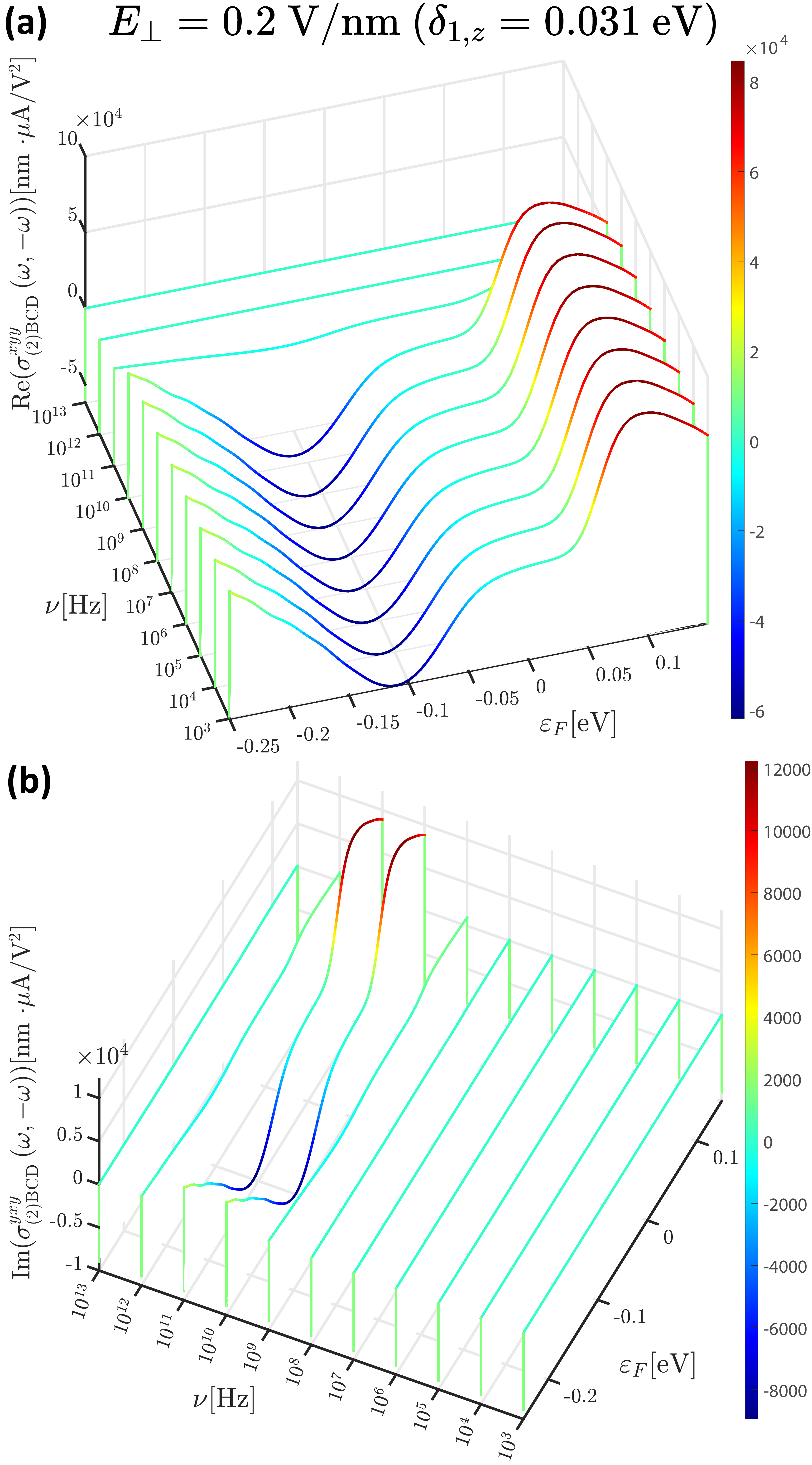}
		\end{center}
		\caption{ (Color online) (a,b) The frequency $ \nu $ and Fermi level $ \varepsilon_F $ dependence of (a) LPGE conductivity $\operatorname{Re}\left(\sigma_{(2) \mathrm{BCD} }^{x y y}(\omega,-\omega)\right)$ and (b) CPGE conductivity $\operatorname{Im}\left(\sigma_{(2) \mathrm{BCD} }^{y x y}(\omega,-\omega)\right)$ for $E_{\perp}=0.2 \mathrm{~V} / \mathrm{nm}$. Here, the frequencies are taken as $10^3$, $10^4$, $10^5$, $\cdots$, $10^{13} \mathrm{~Hz}$. The Fermi level range is from $-0.25$ eV to $0.15$ eV with an interval of 0.005 eV. The relaxation time is taken as 5 ps.} \label{Fig4}
	\end{figure}

	With the use of the traditional BCD formula obtained from the semi-classical Boltzmann equation, the NHE of monolayer WTe$_2$ in the case where $\varepsilon _F$ can be varied and in the case where both $E_{\perp}$ and $\varepsilon _F$ can be varied were studied theoretically by density functional theory (DFT) in Ref.~\onlinecite{You2018} and Ref.~\onlinecite{Zhang2018}, respectively.
	However, only the BCD contribution is included in Refs. \onlinecite{You2018,Zhang2018} at low frequency and there is no band gap in the band dispersion in Ref. \onlinecite{Zhang2018} especially, which may not be proper.
	We first investigate the important effect of the Fermi level on the LPGE or DC current of NHE at 1000 Hz and then compare the results with those in Ref.~\onlinecite{You2018}.
	We consider the following case: the alternating electric field with a low frequency $\nu = 1000 $ $ \text{Hz} $ (angular frequency $ \omega=2\pi \nu $) is set to be along the $ y $ direction \cite{Ma2018,Shi2019}, and the symmetry analysis shows that the linear charge current in the transverse direction will disappear at this time \cite{Sodemann2015}. $ \delta_{1, x} $ and $ \delta_{1, z} $ are set to $0.01$ $ \text{eV}$ and $0.031$ $ \text{eV}$, respectively, which corresponds to $E_{\perp}=0.2 \mathrm{~V} / \mathrm{nm}$. When the Fermi level is $ 0 $ $ \text{eV} $, from the numerical results we find that the only non-vanishing LPGE conductivity is $ \sigma^{xyy}_{(2)}(\omega,-\omega) $, and it is almost all contributed by the BCD term. The real part of $ \sigma^{xyy}_{(2)}(\omega,-\omega) $ is $ -781 $ $ \text{nm} \cdot \mu \text{A}/\text{V}^2 $ and the imaginary part can be neglected. The integrand of the BCD conductivity of LPGE in Eq. (\ref{2nd-conductivity-BCD}) is given as (see SM \cite{supplemental_material})
	\begin{widetext}
		\begin{equation} \begin{aligned}   \label{I_BCD-LPGE-xyy}
				I_{\mathrm{BCD}, \mathrm{LPGE}}^{x y y}(\omega,-\omega) & \equiv \operatorname{Re}\left\{\frac{1}{2}\left[\sum_{n m} \frac{d^\omega}{\hbar} \frac{d_{n m}^0}{\hbar} \varepsilon_{m n} \xi_{m n}^x \xi_{n m}^y \partial_{k_y} f_{m n}(\vec{k})+[(y, \omega) \leftrightarrow(y,-\omega)]\right]\right\}\\
				&=-\frac{2}{\hbar}\frac{\Gamma}{\omega ^2+\Gamma ^2}\left\{ \sum_{nm}{\frac{\left( \varepsilon _{nm}/\hbar \right) ^2}{\left( \varepsilon _{nm}/\hbar \right) ^2+\Gamma ^2}\mathrm{Im}\left( \xi _{mn}^{x}\xi _{nm}^{y} \right) \partial _{k_y}f_n(\vec{k})} \right.\\
				&\quad \left. -\sum_{nm}{\frac{\Gamma \varepsilon _{mn}/\hbar}{\left( \varepsilon _{nm}/\hbar \right) ^2+\Gamma ^2}\mathrm{Re}\left( \xi _{mn}^{x}\xi _{nm}^{y} \right) \partial _{k_y}f_n(\vec{k})} \right\},\\
		\end{aligned}  \end{equation}
	\end{widetext}
	whose $ \vec{k} $-space distribution is plotted in Fig.~\ref{Fig3}.
	$ I_{\mathrm{BCD}, \mathrm{LPGE} }^{x y y} $ are very large near $\Gamma$ point, but show antisymmetric with respect to $\Gamma$ point [shown in Fig.~\ref{Fig3}(a)]. This gives rise to zero BCD-induced LPGE conductivity after integral over the momentum space.
	To reflect the contributions from the states which are far away from $\Gamma$ point to $I_{\mathrm{BCD}, \mathrm{LPGE} }^{x y y}$, we remove the region nearby $\Gamma$ point [which takes a rectangular region near $ \Gamma $ point in Fig.~\ref{Fig3}(b)], but retain the region just around it in Fig.~\ref{Fig3}(b).
	%
	%
	%If we draw the $ \vec{k} $-space distribution of $ I_{\mathrm{BCD}, \mathrm{LPGE} }^{x y y} $ after removing the large magnitude of $ I_{\mathrm{BCD}, \mathrm{LPGE}}^{x y y} $ in the rectangular region near $ \Gamma $ point [see Fig.~\ref{Fig3}(b)],
	%
	Before we start our discussions on the behavior of $I_{\mathrm{BCD}, \mathrm{LPGE} }^{x y y}$ in the momentum space, we have to emphasize that $I_{\mathrm{BCD}, \mathrm{LPGE} }^{x y y}$ is generally not the same to the usual Berry curvature dipole, i.e., $\sum_n{f_n(\vec{k})}\partial _{k_y}\Omega _{n}^{z} $.
	$I_{\mathrm{BCD}, \mathrm{LPGE} }^{x y y}$ is only reduced to the usual Berry curvature dipole at the limit of $\Gamma \ll\left|\varepsilon_{n m}/\hbar\right| $ (see Section \ref{discussion}).
	Thus it loses a clear meaning of Berry curvature dipole, but we can see that the distribution of integrand in momentum space shows positive-negative distributions along an axis, which is analogous to a ``dipole'' in momentum space. For brevity, we call this as dipole distribution.
	It can be seen that there are two centers of these dipole distributions [at $(k_x, k_y)=(0, \pm0.105)$ $ \text{\AA}^{-1} $], which give rise to the main contribution to the BCD-induced LPGE conductivity.
	These dipole distributions are a subtle consequence of the product of $ d^{\omega} d_{n m}^{0} \varepsilon_{m n} $, $ \partial_{k_y} f_{m n} $ and $ \xi_{m n}^x \xi_{n m}^y $ in Eq.~(\ref{I_BCD-LPGE-xyy}).
	For a comparison, if the Fermi level is set to be $ 0.095 $ $ \text{eV} $, which corresponds to gap-opening points in the energy band dispersion located at $Q$ and $Q^{\prime}$ in momentum space [see Figs.~\ref{Fig2}(a) and \ref{Fig1}(d)]. It also displays dipole-distribution features near these gap-opening points $Q$ and $Q^{\prime} $ [see Fig.~\ref{Fig3}(c)].
	The real part of $ \sigma^{xyy}_{(2)}(\omega,-\omega) $ tremendously increases to $ 8.582\times 10^{4} $ $ \text{nm} \cdot \mu \text{A}/\text{V}^2 $ (from $ -781$ nm$\cdot \mu\text{A}/\text{V}^2 $ for Fermi level at 0 eV), which is mainly contributed by these dipoles and the order of magnitude is consistent with $ \sigma^{xyy}_{(2)}(\omega,-\omega) $ estimated in Ref.~\onlinecite{You2018} \cite{compare_order}.
	Furthermore, when $E_{\perp}=0.2 \mathrm{~V} / \mathrm{nm}$ and $ \nu$ = 1000 Hz, the dependence of the LPGE or DC current of NHE on the Fermi level can be found in Fig.~\ref{Fig4}(a), and its trend and order of magnitude are qualitatively consistent with Ref.~\onlinecite{You2018} (see Fig.~2(b) of Ref.~\onlinecite{You2018}).
	
	%%%%%%%%%%%%%%%%%%%%%%%%%%%%%%%%%%%%%%%%%%%%%%%%%%%%%%%%%%%%%%%%%%%%%%%%%%%%%%%%%%%%%%%%%%%%%%%%%%%%%%%%%%%%%%%%%%%%%%%%%%%%%%%%%%%%%%%%%
	%%%%%%%%%%%%%%%%%%%%%%%%%%%%%%%%%%%%%%%%%%%%%%%%%%%%%%%%%%%%%%%%%%%%%%%%%%%%%%%%%%%%%%%%%%%%%%%%%%%%%%%%%%%%%%%%%%%%%%%%%%%%%%%%%%%%%%%%%

	%%%%%%%%%%%%%%%%%%%%%%%%%%%%%%%%%%%%%%%%%%%%%%%%%%%%%%%%%%%%%%%%%%%%%%%%%%%%%%%%%%%%%%%%%%%%%%%%%%%%%%%%%%%%%%%%%%%%%%%%%%%%%%%%%%%%%%%%%
	\begin{figure*}[t!]
		\begin{center}
			\includegraphics[width=2\columnwidth]{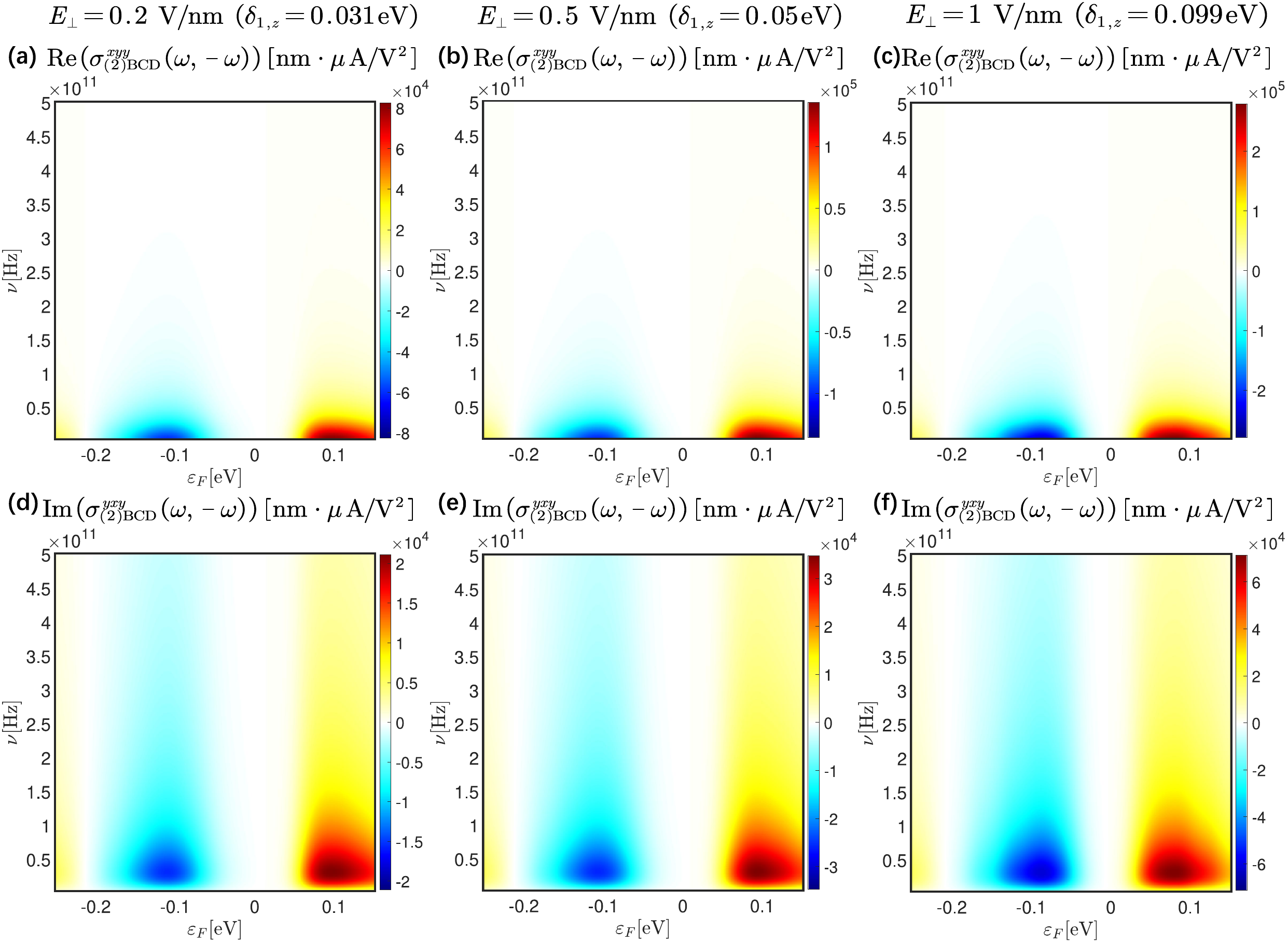}
		\end{center}
		\caption{ (Color online) (a,b,c) The frequency $ \nu $ and Fermi level $ \varepsilon_F $ dependence of LPGE conductivity $\operatorname{Re}\left(\sigma_{(2) \mathrm{BCD} }^{x y y}(\omega,-\omega)\right)$ for (a) $E_{\perp}=0.2 \mathrm{~V} / \mathrm{nm}$, (b) $E_{\perp}=0.5 \mathrm{~V} / \mathrm{nm}$ and (c) $E_{\perp}=1 \mathrm{~V} / \mathrm{nm}$. (d,e,f) The frequency $ \nu $ and Fermi level $ \varepsilon_F $ dependence of CPGE conductivity $\operatorname{Im}\left(\sigma_{(2) \mathrm{BCD} }^{y x y}(\omega,-\omega)\right)$ for (d) $E_{\perp}=0.2 \mathrm{~V} / \mathrm{nm}$, (e) $E_{\perp}=0.5 \mathrm{~V} / \mathrm{nm}$ and (f) $E_{\perp}=1 \mathrm{~V} / \mathrm{nm}$. Here, the frequency range for numerical calculations is from 5 GHz to 500 GHz with an interval of 5 GHz. The Fermi level range is from $-0.25$ eV to $0.15$ eV with an interval of 0.005 eV. The relaxation time is taken as 5 ps.} \label{Fig5}
	\end{figure*}  %Use the figure* environment to get a wide figure that spans the page in \texttt{twocolumn} formatting
	%%%%%%%%%%%%%%%%%%%%%%%%%%%%%%%%%%%%%%%%%%%%%%%%%%%%%%%%%%%%%%%%%%%%%%%%%%%%%%%%%%%%%%%%%%%%%%%%%%%%%%%%%%%%%%%%%%%%%%%%%%%%%%%%%%%%%%%%%
	
	\subsection{LPGE and CPGE at higher frequencies}
	Now we turn to the case of higher frequencies. For LPGE with the alternating electric field along the $ y $ direction, from Eq.~(\ref{linearly_polarized-PGE}) we know that only $\operatorname{Re}\left(\sigma_{(2)}^{x y y}( \omega,-\omega)\right)$ and $\operatorname{Re}\left(\sigma_{(2)}^{y y y}( \omega,-\omega)\right)$ need to be calculated. For CPGE, from Eq.~(\ref{CPGE_response}) we know that only $\operatorname{Im}\left(\sigma_{(2)}^{x x y}( \omega,-\omega)\right)$ and $\operatorname{Im}\left(\sigma_{(2)}^{y x y}( \omega,-\omega)\right)$ need to be calculated.
	We first calculated the LPGE and CPGE conductivities with a large range of frequencies and Fermi levels, and the results show that in the range of $10^3$ Hz-$10^{13}$ Hz, $\operatorname{Re}\left(\sigma_{(2)}^{y y y}(\omega,-\omega)\right)$ and $\operatorname{Im}\left(\sigma_{(2)}^{x x y}( \omega,-\omega)\right)$ vanish while $\operatorname{Re}\left(\sigma_{(2)}^{x y y}(\omega,-\omega)\right)$ and $\operatorname{Im}\left(\sigma_{(2)}^{y x y}( \omega,-\omega)\right)$ are present, as well as they are almost all contributed by the $\operatorname{Re}\left(\sigma_{(2)\mathrm{BCD}}^{x y y}(\omega,-\omega)\right)$ and $\operatorname{Im}\left(\sigma_{(2)\mathrm{BCD}}^{y x y}( \omega,-\omega)\right)$, respectively. The frequency $ \nu $ and Fermi level $ \varepsilon_F $ dependence of $\operatorname{Re}\left(\sigma_{(2) \mathrm{BCD} }^{x y y}(\omega,-\omega)\right)$ and $\operatorname{Im}\left(\sigma_{(2) \mathrm{BCD} }^{y x y}(\omega,-\omega)\right)$ for $E_{\perp}=0.2 \mathrm{~V} / \mathrm{nm}$ are shown in Fig.~\ref{Fig4} and our calculations also find the same trend as Fig.~\ref{Fig4} when $E_{\perp}$ increases to 0.5 V/nm and 1 V/nm, except that the value of conductivities increases. From the aspect of Fermi level, the maximum value of PGE conductivities occurs when the Fermi level is equal to $ 0.095 $ eV and $ -0.11 $ eV, which correspond to the vicinity of the gap-opening points in the band dispersion [see Fig.~\ref{Fig2}(a)]. The value and sign of the PGE conductivities can be changed greatly when changing the Fermi level through the gate voltage, which may be useful for fabricating electrically switchable rectifiers \cite{Zhou2020}. From the aspect of frequency, in the radio region (about less than $10^9$ Hz), the LPGE varies very little with frequency and the CPGE disappears. When $ \nu $ is greater than $10^9$ Hz, the LPGE conductivity and the CPGE conductivity start to gradually decrease and increase, respectively, with increasing frequency. Thus, we next take a careful look at LPGE and CPGE in the region of 5 GHz-500 GHz, which covers a large part of the microwave region (0.3 GHz-300 GHz). Figs.~\ref{Fig5}(a)-(c) show the rapid decrease of $\operatorname{Re}\left(\sigma_{(2) \mathrm{BCD}}^{x y y}(\omega,-\omega)\right) $ as the frequency increases from 5 GHz to 100 GHz. The frequency dependence of the LPGE and CPGE in Figs.~\ref{Fig4} and \ref{Fig5} can be directly explained by $\operatorname{Re}\left(\sigma_{(2) \mathrm{BCD}}^{x y y}(\omega,-\omega)\right) \propto 1/(\omega^ 2+\Gamma^2)$ and $\operatorname{Im}\left(\sigma_{(2) \mathrm{BCD}}^{yx y }(\omega,-\omega)\right) \propto \omega/(\omega^ 2+\Gamma^2)$, respectively, which can be known from Eqs.~(\ref{I_BCD-LPGE-xyy}) and (\ref{I_BCD-CPGE-yxy}). Among Figs.~\ref{Fig5}(d)-\ref{Fig5}(f), the maximum CPGE response with a value of $7.14\times 10^4$ $ \text{nm} \cdot \mu \text{A}/\text{V}^2 $ occurs when $ \nu$ = 30 GHz, $\varepsilon _F=0.08 \mathrm{~eV}$ and $E_{\perp}=1 \mathrm{~V} / \mathrm{nm}$.

	%%%%%%%%%%%%%%%%%%%%%%%%%%%%%%%%%%%%%%%%%%%%%%%%%%%%%%%%%%%%%%%%%%%%%%%%%%%%%%%%%%%%%%%%%%%%%%%%%%%%%%%%%%%%%%%%%%%%%%%%%%%%%%%%%%%%%%%%%
	\begin{figure*}[t!]
		\begin{center}
			\includegraphics[width=2\columnwidth]{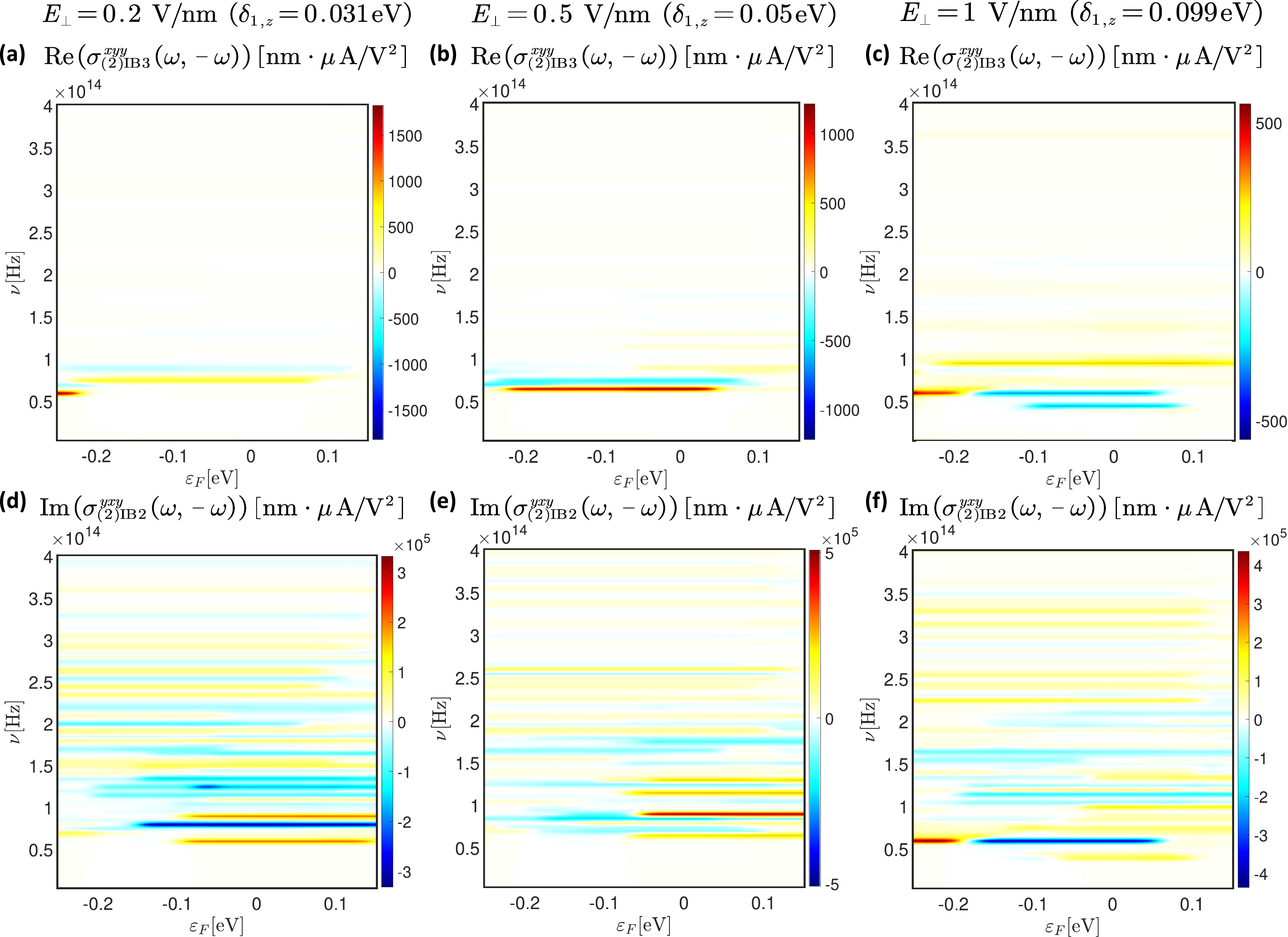}
		\end{center}
		\caption{ (Color online) (a,b,c) The frequency $ \nu $ and Fermi level $ \varepsilon_F $ dependence of LPGE conductivity $\operatorname{Re}\left(\sigma_{(2) \mathrm{IB} 3}^{x y y}(\omega,-\omega)\right)$ for (a) $E_{\perp}=0.2 \mathrm{~V} / \mathrm{nm}$, (b) $E_{\perp}=0.5 \mathrm{~V} / \mathrm{nm}$ and (c) $E_{\perp}=1 \mathrm{~V} / \mathrm{nm}$. (d,e,f) The frequency $ \nu $ and Fermi level $ \varepsilon_F $ dependence of CPGE conductivity $\operatorname{Im}\left(\sigma_{(2) \mathrm{IB} 2}^{y x y}(\omega,-\omega)\right)$ for (d) $E_{\perp}=0.2 \mathrm{~V} / \mathrm{nm}$, (e) $E_{\perp}=0.5 \mathrm{~V} / \mathrm{nm}$ and (f) $E_{\perp}=1 \mathrm{~V} / \mathrm{nm}$. Here, the frequency range for numerical calculations is from 5 THz to 400 THz with an interval of 5 THz. The Fermi level range is from $-0.25$ eV to $0.15$ eV with an interval of 0.005 eV. The relaxation time is taken as 5 ps.} \label{Fig6}
	\end{figure*}  %Use the figure* environment to get a wide figure that spans the page in \texttt{twocolumn} formatting
	%%%%%%%%%%%%%%%%%%%%%%%%%%%%%%%%%%%%%%%%%%%%%%%%%%%%%%%%%%%%%%%%%%%%%%%%%%%%%%%%%%%%%%%%%%%%%%%%%%%%%%%%%%%%%%%%%%%%%%%%%%%%%%%%%%%%%%%%%

	Many nonlinear optics experiments are measured in the infrared region of light (0.3 THz-400 THz) \cite{Xu2018,Sie2019,Drueke2021,Aoki2022}. When the frequency $ \nu $ is greater than 400 THz, the corresponding photon energy is greater than 1.65 eV. As can be seen from the band dispersion calculated by DFT in Fig.~1(e) of Ref.~\onlinecite{Xu2018}, more bands will be involved in the interband transitions at this time, and our six-band model may be inaccurate.
	Therefore, we next choose the frequency range 5 THz-400 THz to investigate the dependence of LPGE and CPGE on Fermi level and frequency under different $E_{\perp}$. For LPGE with alternating electric field along the $ y $-direction, the non-vanishing conductivity is $ \operatorname{Re}\left(\sigma^{xyy}_{(2)}(\omega,-\omega)\right) $, which is almost all contributed by the IB3 term.
	As shown in Figs.~\ref{Fig6}(a)-\ref{Fig6}(c), $\operatorname{Re}\left(\sigma_{(2) \mathrm{IB} 3}^{x y y}(\omega,-\omega)\right)$ has relatively large values of about 500 $ \text{nm} \cdot \mu \text{A}/\text{V}^2 $ to 1000 $ \text{nm} \cdot \mu \text{A}/\text{V}^2 $ in the range of 50 THz-100 THz.
	For CPGE, $\operatorname{Im}\left(\sigma_{(2)}^{x x y}(\omega,-\omega)\right)$ disappears while $\operatorname{Im}\left(\sigma_{(2)}^{y x y}(\omega,-\omega)\right)$ is present. $\operatorname{Im}\left(\sigma_{(2)}^{y x y}(\omega,-\omega)\right)$ is mainly contributed by the IB2 term and the contributions from the BCD and IB3 terms are smaller than that from the IB2 term by an order of $ 10^3 $.
	Among Figs.~\ref{Fig6}(d)-\ref{Fig6}(f), the maximum CPGE response with a value of $5.07\times 10^5$ $ \text{nm} \cdot \mu \text{A}/\text{V}^2 $ occurs when $ \nu$ = 90 THz, $\varepsilon _F=0.07 \mathrm{~eV}$ and $E_{\perp}=0.5 \mathrm{~V} / \mathrm{nm}$. In a word, the $ \nu $, $ \varepsilon_F $ and $ E_{\perp} $ dependence of PGE are complicated in the infrared region.

	It can be known directly from Eq.~(S64) of SM \cite{supplemental_material} that our theory of the independent particle approximation only considers transitions connecting identical $\vec{k}$ points \cite{Haug2004}.
	Therefore, we plot the energy difference $\Delta \varepsilon \left(k_x, k_y\right)$ between the lowest conduction band and the highest valence band (see Fig.~S1 of SM \cite{supplemental_material}) and find that the minimum of $\Delta \varepsilon \left(k_x, k_y\right)$ is located near the gap-opening points.
	When $E_{\perp}$ = 0.2 V/nm, 0.5 V/nm and 1 V/nm, the minimum values of $\Delta \varepsilon \left(k_x, k_y\right)$ are 0.233 eV, 0.2065 eV and 0.133 eV respectively  (consistent with previous calculations in Ref.~\onlinecite{Xu2018}), which correspond to the required frequencies of 56.34 THz, 49.93 THz and 32.16 THz for the interband transitions.
	In the infrared region, large values of PGE exist only when the frequency is greater than 50 THz (see Fig.~\ref{Fig6}), which implies a possible connection between the IB2, IB3 terms and the interband transitions.

	\subsection{LSHG (or Second harmonic current of NHE) and CSHG}
\begin{figure}[t!]
		\begin{center}
			\includegraphics[width=1\columnwidth]{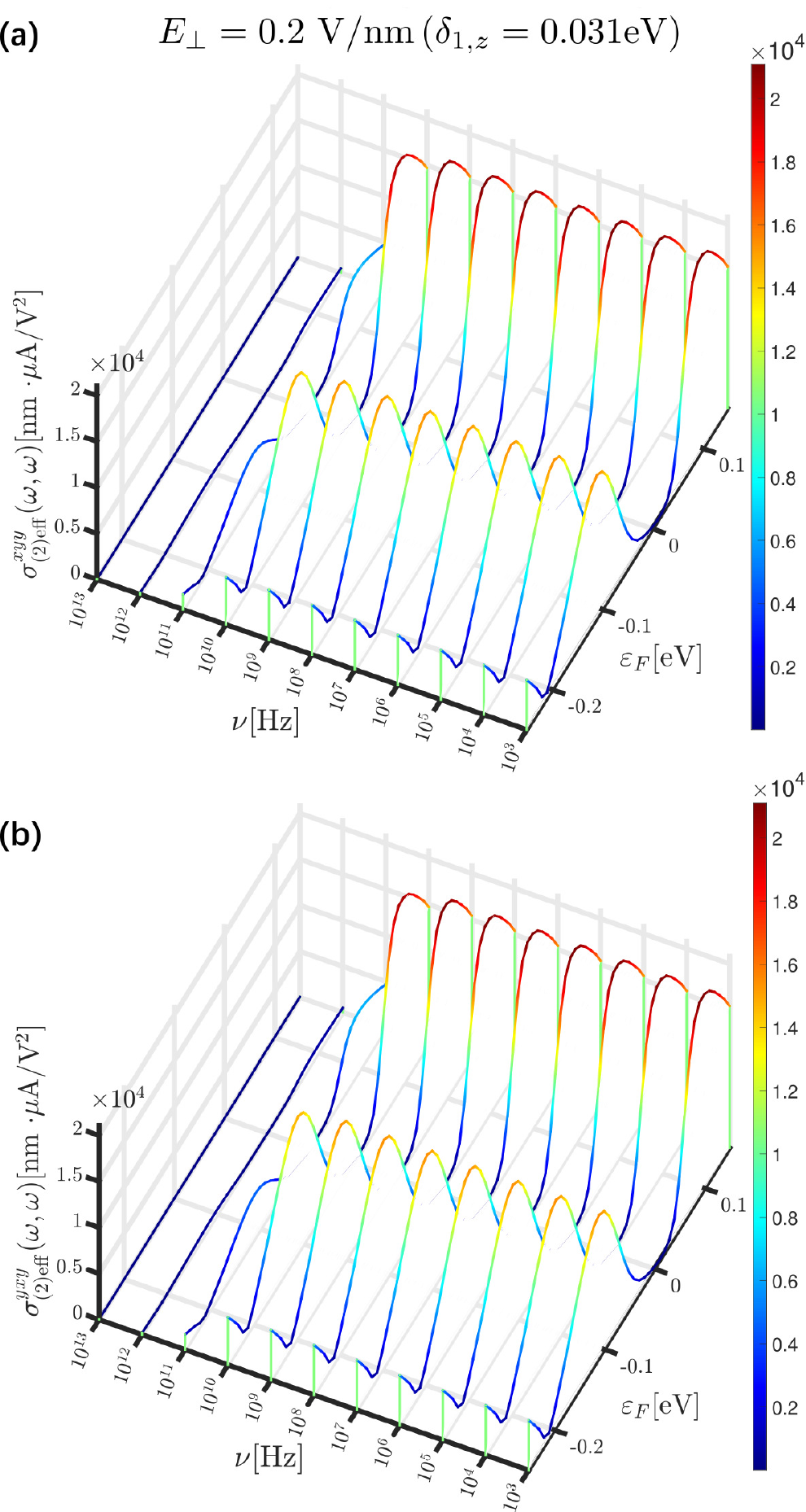}
		\end{center}
		\caption{ (Color online) (a,b) The frequency $ \nu $ and Fermi level $ \varepsilon_F $ dependence of (a) LSHG conductivity $\sigma_{(2) \mathrm{eff}}^{x y y}(\omega, \omega)$ and (b) CSHG conductivity $\sigma_{(2) \mathrm{eff}}^{y x y}(\omega, \omega)$ for $E_{\perp}=0.2 \mathrm{~V} / \mathrm{nm}$. Here, the frequencies are taken as $10^3$, $10^4$, $10^5$, $\cdots$, $10^{13} \mathrm{~Hz}$. The Fermi level range is from $-0.25$ eV to $0.15$ eV with an interval of 0.01 eV. The relaxation time is taken as 5 ps.} \label{Fig7}
	\end{figure}

	For LSHG with the alternating electric field along the $ y $ direction, from Eq.~(\ref{LSHG-eff}) we know that only $\sigma_{(2) \mathrm{eff}}^{x y y}(\omega, \omega)$ and $\sigma_{(2) \mathrm{eff}}^{y y y}(\omega, \omega)$ need to be calculated. For CSHG, from Eq.~(\ref{CSHG-eff}) we know that only $\sigma_{(2) \mathrm{eff}}^{x x y}(\omega, \omega)$ and $\sigma_{(2) \mathrm{eff}}^{y x y}(\omega, \omega)$ need to be calculated.
	We first calculated the effective LSHG and CSHG conductivities with a large range of frequencies and Fermi levels by using Eqs.~(\ref{2nd-conductivity-another}), (\ref{LSHG-eff}) and (\ref{CSHG-eff}), and the results show that in the range of $10^3$ Hz-$10^{13}$ Hz, $\sigma_{(2) \mathrm{eff}}^{y y y}(\omega, \omega)$ and $\sigma_{(2) \mathrm{eff}}^{x x y}(\omega, \omega)$ vanish while $\sigma_{(2) \mathrm{eff}}^{x y y}(\omega, \omega)$ and $\sigma_{(2) \mathrm{eff}}^{y x y}(\omega, \omega)$ are present, as well as they are almost all contributed by the real and imaginary parts of the BCD term when $\nu$ is less than $10^{12}$ Hz. The frequency $ \nu $ and Fermi level $ \varepsilon_F $ dependence of $\sigma_{(2) \mathrm{eff}}^{x y y}(\omega, \omega)$ and $\sigma_{(2) \mathrm{eff}}^{y x y}(\omega, \omega)$ for $E_{\perp}=0.2 \mathrm{~V} / \mathrm{nm}$ are shown in Fig.~\ref{Fig7} and our calculations also find the same trend as Fig.~\ref{Fig7} when $E_{\perp}$ increases to 0.5 V/nm and 1 V/nm, except that the value of conductivities increases. Surprisingly, the $ \nu $ and $ \varepsilon_F $ dependence of CSHG and LSHG have almost the same trend and values by comparing Fig.~\ref{Fig7}(a) and Fig.~\ref{Fig7}(b). From the aspect of Fermi level, the two large peaks of SHG conductivities occur when the Fermi level is equal to the energy corresponding to the vicinity of the gap-opening points in the band dispersion. From the aspect of frequency, in the radio region (about less than $10^9$ Hz), the SHG conductivities are almost constant with increasing frequency. When $ \nu $ is greater than $10^9$ Hz, the SHG conductivities start to gradually decrease with increasing frequency.

\begin{figure*}[t!]
		\begin{center}
			\includegraphics[width=2\columnwidth]{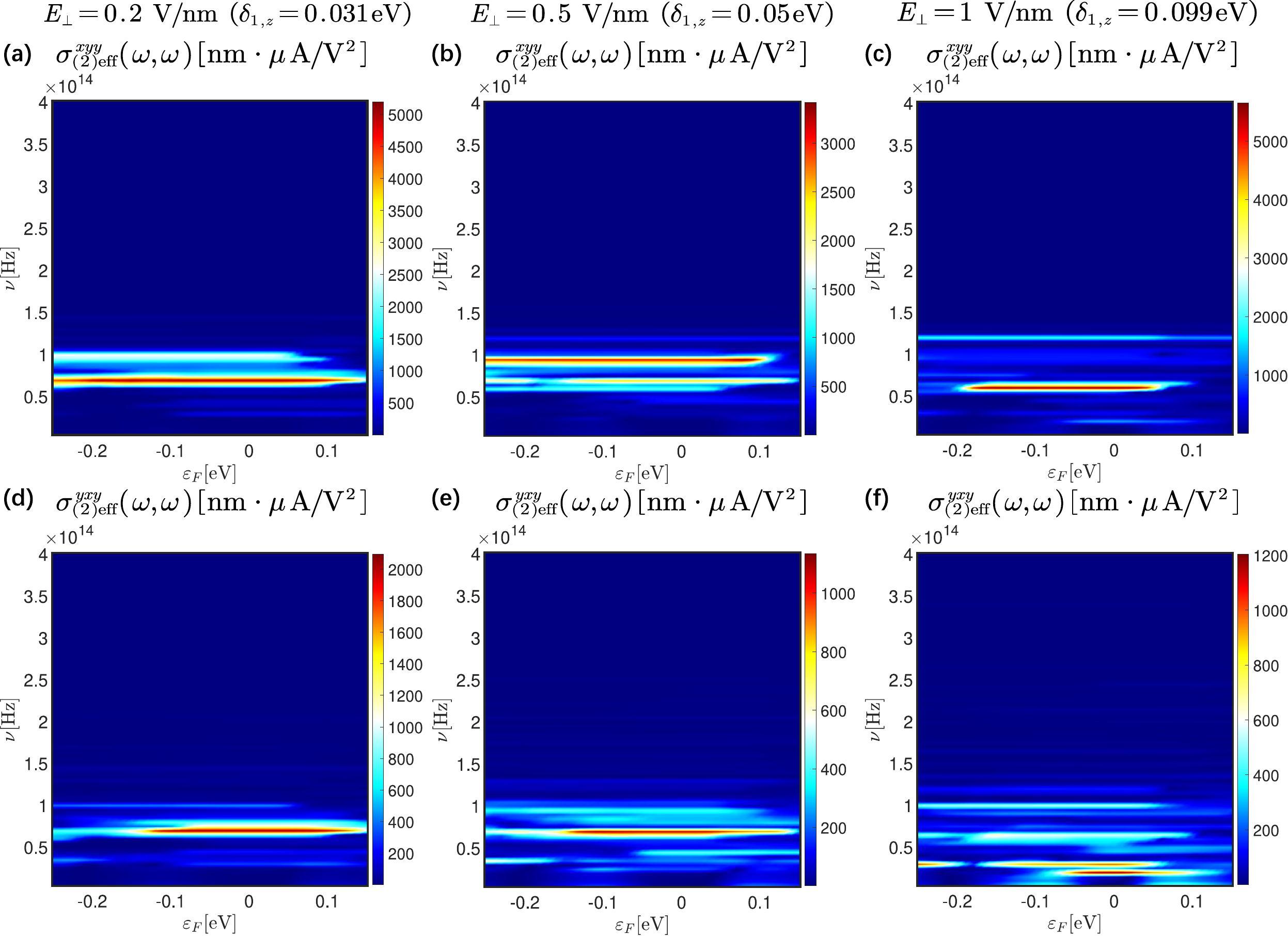}
		\end{center}
		\caption{ (Color online) (a,b,c) The frequency $ \nu $ and Fermi level $ \varepsilon_F $ dependence of LSHG conductivity $\sigma_{(2) \mathrm{eff}}^{x y y}(\omega, \omega)$ for (a) $E_{\perp}=0.2 \mathrm{~V} / \mathrm{nm}$, (b) $E_{\perp}=0.5 \mathrm{~V} / \mathrm{nm}$ and (c) $E_{\perp}=1 \mathrm{~V} / \mathrm{nm}$. (d,e,f) The frequency $ \nu $ and Fermi level $ \varepsilon_F $ dependence of CSHG conductivity $\sigma_{(2) \mathrm{eff}}^{y x y}(\omega, \omega)$ for (d) $E_{\perp}=0.2 \mathrm{~V} / \mathrm{nm}$, (e) $E_{\perp}=0.5 \mathrm{~V} / \mathrm{nm}$ and (f) $E_{\perp}=1 \mathrm{~V} / \mathrm{nm}$. Here, the frequency range for numerical calculations is from 5 THz to 400 THz with an interval of 5 THz. The Fermi level range is from $-0.25$ eV to $0.15$ eV with an interval of 0.01 eV. The relaxation time is taken as 5 ps.} \label{Fig8}
	\end{figure*}

	Next we turn to the infrared region (0.3 THz-400 THz, i.e., 0.3$\times$10$^{12}$ Hz-400$\times$10$^{12}$ Hz). It remains that $\sigma_{(2) \mathrm{eff}}^{y y y}(\omega, \omega)$ and $\sigma_{(2) \mathrm{eff}}^{x x y}(\omega, \omega)$ vanish while $\sigma_{(2) \mathrm{eff}}^{x y y}(\omega, \omega)$ and $\sigma_{(2) \mathrm{eff}}^{y x y}(\omega, \omega)$ exist.
	As shown in Fig.~\ref{Fig8}, the $ \nu $, $ \varepsilon_F $ and $ E_{\perp} $ dependence of the SHG are complicated and there are contributions of about the same order of magnitude from the BCD and IB3 terms. The SHG conductivities have relatively large values in the range of 5 THz-125 THz. Comparing Figs.~\ref{Fig6}(d)-(f) and Figs.~\ref{Fig8}(d)-(f), we can know from Eqs.~(\ref{LC-RC-L-0w-wte2}) and (\ref{LC-RC-2w-wte2}) that in the frequency range of 125 THz-300 THz and in the $y$-direction, the CSHG almost disappears and only a large CPGE exists, which is a very special property that may be attractive for circularly polarized infrared photodetection \cite{Zhang2021,Wei2022} and electromagnetic wave energy harvesting rectifiers \cite{Zhou2020} without the disturbance of CSHG.

	\subsection{The case of relaxation time of 10 fs}
	\par Relaxation times of 10 fs have also been measured in some experimental samples \cite{Aoki2022,Ma2018,Xu2018,Qin2021}. Therefore we next investigate the case of $\tau$ = 10 fs. In the range of $10^3$ Hz-$10^{14}$ Hz, for LPGE, only $\operatorname{Re}\left(\sigma_{(2)}^{x y y}(\omega,-\omega)\right)$ exists, which is mainly contributed by the BCD term. The $\nu$ and $\varepsilon_F$ dependence of $\operatorname{Re}\left(\sigma_{(2) \mathrm{BCD}}^{x y y}(\omega,-\omega)\right)$ is also similar to Fig.~\ref{Fig4}(a) (see Fig.~\ref{Fig9}). Compared to the case of $\tau=5$ ps, the difference is that the order of magnitude of $\operatorname{Re}\left(\sigma_{(2) \mathrm{BCD}}^{x y y}(\omega,-\omega)\right)$ decreases by a factor of 1000, and from the viewpoint of frequency, the LPGE photocurrent only begins to decrease gradually when the $\nu$ is greater than $10^{12}$ Hz. For CPGE, we find that there is almost no CPGE photocurrent when $\nu$ is less than $10^{13}$ Hz. As shown in Fig.~\ref{Fig10}, in the region of 5 THz-400 THz [it covers a large part of the infrared region (0.3 THz-400 THz)], the LPGE current still exists in the range of 5 THz-100 THz, which is mainly contributed by the BCD term. For CPGE, BCD, IB2, and IB3 terms all have about the same order of magnitude contribution to the total conductivity, so the situation becomes more complicated than the case of $\tau=5$ ps.
	\par The CPGE in the WTe$_2$ monolayer at $ \varepsilon_F\approx0 ~\text{eV}$, $ \tau = 10$ fs, $ \nu = 29$ THz and different $ E_{\perp} $ has been investigated experimentally and theoretically in Ref.~\onlinecite{Xu2018}. Our theoretical results for $\operatorname{Im}\left(\sigma_{(2)}^{y x y}(\omega,-\omega)\right)$ in Fig.~\ref{Fig10}(e) are consistent in the order of magnitude with the experimentally measured approximate value 177 $ \text{nm} \cdot \mu \text{A}/\text{V}^2 $ \cite{Xu2018} when $ E_{\perp} = 0.5$ V/nm.

	\section{Discussion}\label{discussion}
	
	In the frequency range we considered above, the wavelength of the applied light is much larger than the characteristic size of the system \cite{Zhumagulov2022,Yu2010,Ma2021}, the wave vector of the photon is small and the photon drag effect \cite{Ganichev2006,Ma2021} can be ignored. The experimental scheme for \textcolor{red}{measuring nonlinear optical conductivities from radio to infrared region is illustrated in the SM \cite{supplemental_material} and the schematic illustration of photocurrents that need to be measured in order to verify our theory in $ T_d-\text{WTe}_2 $ monolayer is shown in Fig.~\ref{Fig11}.}
	
	\par
\begin{figure*}[t!]
		\begin{center}
			\includegraphics[width=2\columnwidth]{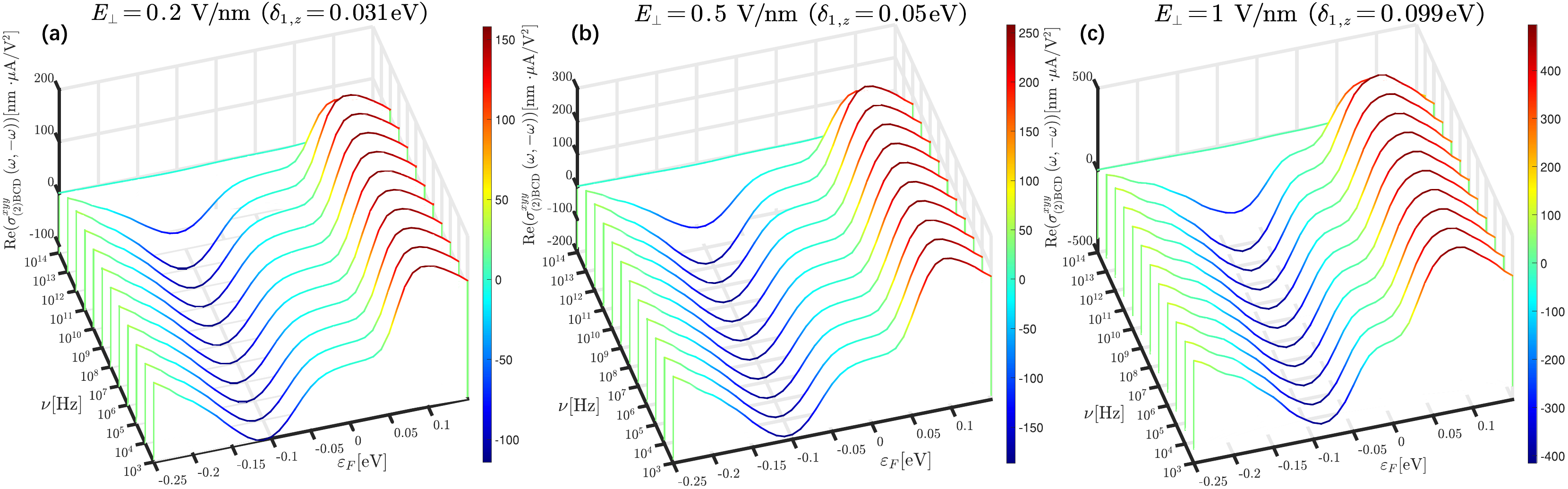}
		\end{center}
		\caption{ (Color online) (a,b,c) The frequency $ \nu $ and Fermi level $ \varepsilon_F $ dependence of LPGE conductivity $\operatorname{Re}\left(\sigma_{(2) \mathrm{BCD} }^{x y y}(\omega,-\omega)\right)$ for (a) $E_{\perp}=0.2 \mathrm{~V} / \mathrm{nm}$, (b) $E_{\perp}=0.5 \mathrm{~V} / \mathrm{nm}$ and (c) $E_{\perp}=1 \mathrm{~V} / \mathrm{nm}$. Here, the frequencies are taken as $10^3$, $10^4$, $10^5$, $\cdots$, $10^{14} \mathrm{~Hz}$. The Fermi level range is from $-0.25$ eV to $0.15$ eV with an interval of 0.01 eV. The relaxation time is taken as 10 fs.} \label{Fig9}
	\end{figure*}

\begin{figure*}[t!]
		\begin{center}
			\includegraphics[width=2\columnwidth]{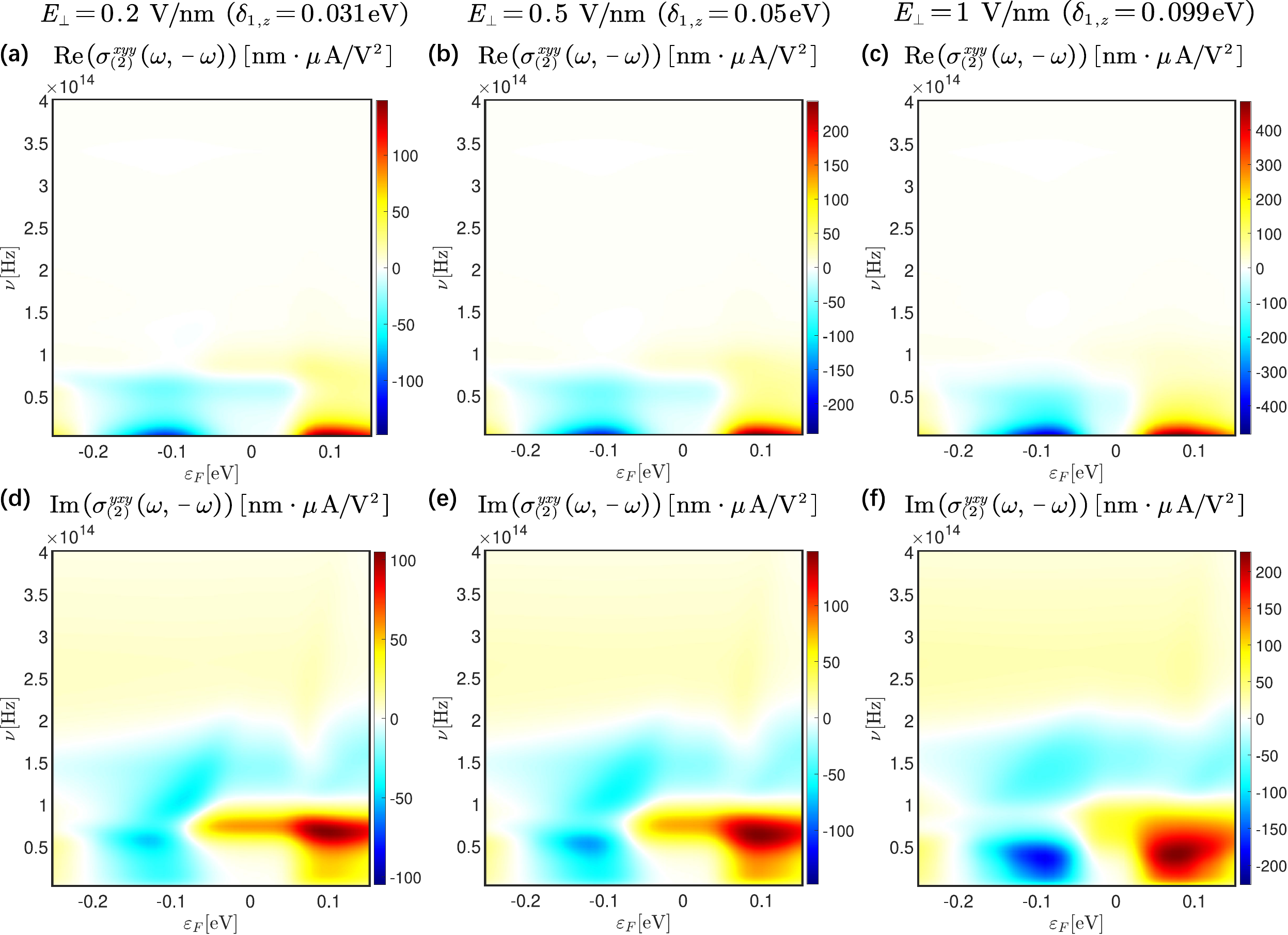}
		\end{center}
		\caption{ (Color online) (a,b,c) The frequency $ \nu $ and Fermi level $ \varepsilon_F $ dependence of LPGE conductivity $\operatorname{Re}\left(\sigma_{(2) }^{x y y}(\omega,-\omega)\right)$ for (a) $E_{\perp}=0.2 \mathrm{~V} / \mathrm{nm}$, (b) $E_{\perp}=0.5 \mathrm{~V} / \mathrm{nm}$ and (c) $E_{\perp}=1 \mathrm{~V} / \mathrm{nm}$. (d,e,f) The frequency $ \nu $ and Fermi level $ \varepsilon_F $ dependence of CPGE conductivity $\operatorname{Im}\left(\sigma_{(2)}^{y x y}(\omega,-\omega)\right)$ for (d) $E_{\perp}=0.2 \mathrm{~V} / \mathrm{nm}$, (e) $E_{\perp}=0.5 \mathrm{~V} / \mathrm{nm}$ and (f) $E_{\perp}=1 \mathrm{~V} / \mathrm{nm}$. Here, the frequency range is from 5 THz to 400 THz with an interval of 5 THz. The Fermi level range is from $-0.25$ eV to $0.15$ eV with an interval of 0.01 eV. The relaxation time is taken as 10 fs.} \label{Fig10}
	\end{figure*}

\begin{figure*}[t!]
		\begin{center}
			\includegraphics[width=1.9\columnwidth]{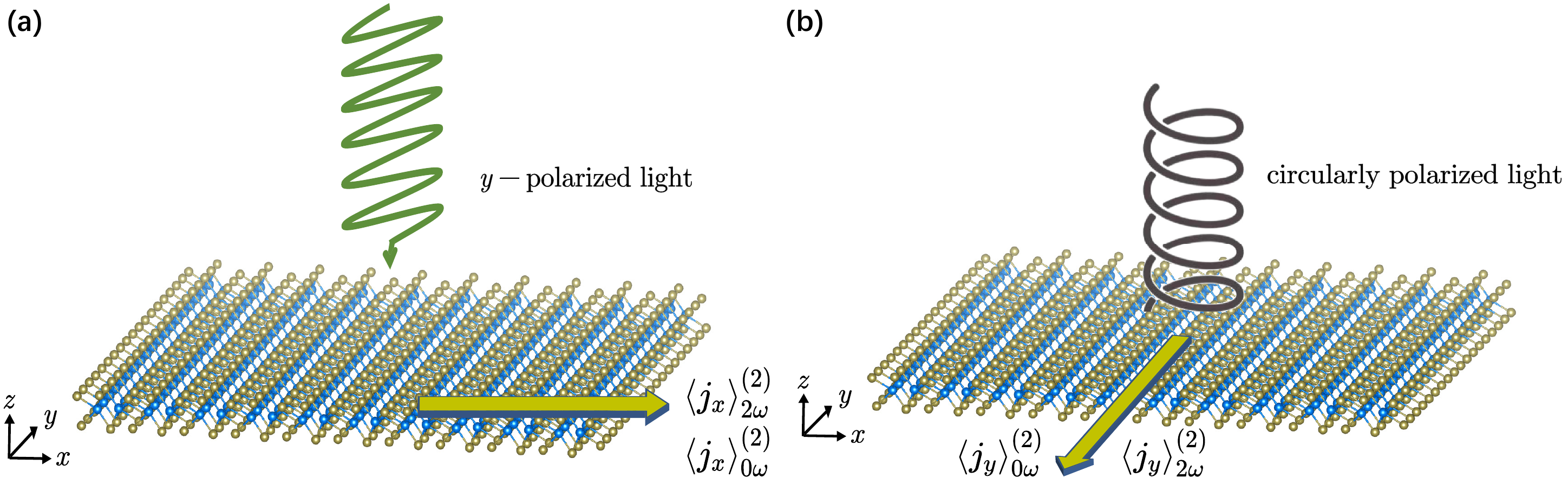}
		\end{center}
		\caption{ (Color online) (a,b) Schematic illustration of \textcolor{red}{photocurrents that need to be measured in order to verify our theory in $ T_d-\text{WTe}_2 $ monolayer.} (a) LPGE and LSHG. A monochromatic linearly polarized light with electric field vector along the $y $-axis incident along the normal to $ T_d-\text{WTe}_2 $ monolayer with $ z=0 $ in the $ xy $ plane. Due to the mirror symmetry $ \mathcal{M}_y $, the second-order currents in the $ y $-direction disappear and only second-order DC and second harmonic currents exist in the $ x $-direction. The second-order currents in the $ x $-direction are induced by the non-zero $\operatorname{Re}\left(\sigma_{(2)}^{x y y}(\omega,-\omega)\right)$ and $\sigma_{(2) \mathrm{eff}}^{x y y}(\omega, \omega)$. (b) CPGE and CSHG. A monochromatic circularly polarized light incident along the normal to $ T_d-\text{WTe}_2 $ monolayer with $ z=0 $ in the $ xy $ plane. From Eqs.~(\ref{LC-RC-L-0w-wte2}) and (\ref{LC-RC-2w-wte2}), we know that DC and second harmonic currents exist in both $ x $ and $ y $ directions under circularly polarized light. However, due to the mirror symmetry $ \mathcal{M}_y $, when the circularly polarized light changes from LC to RC, the DC and second harmonic currents in $ y $ direction will be reversed while the DC and second harmonic currents in $ x $ direction keeps the same magnitude and direction. Thus the CPGE and CSHG responses [see Eqs.~(\ref{CPGE_response}) and (\ref{CSHG-current})] exist only in the $ y $ direction, which are induced by the non-zero $\operatorname{Im}\left(\sigma_{(2)}^{y x y}(\omega,-\omega)\right)$ and $\sigma_{(2) \mathrm{eff}}^{y x y}(\omega, \omega)$.} \label{Fig11}
	\end{figure*}

 In the clean limit, for the PGE, the BCD term exists only under circularly polarized light \cite{Watanabe2021} because the BCD term is purely imaginary [see Eq.~(\ref{traditional_BCD_formula})]. However, Figs.~\ref{Fig4}(a) and \ref{Fig9} show that the BCD term in the case of considering scattering effects has a real part and it contributes significantly to the LPGE, which shows the necessity of considering a finite $\Gamma$ in the calculation of second-order nonlinear optical responses.
	\par Next we discuss the difference between the BCD formula Eq.~(\ref{2nd-conductivity-BCD}) obtained by the quantum kinetics and the traditional BCD formula obtained by the semi-classical Boltzmann equation. According to Eq.~(\ref{2nd-conductivity-BCD}), the integrand of the BCD conductivity of CPGE is given as (see SM \cite{supplemental_material})
	\begin{widetext}
		\begin{equation} \begin{aligned}   \label{I_BCD-CPGE-yxy}
				I_{\mathrm{BCD},\mathrm{CPGE}}^{yxy}(\omega ,-\omega )&\equiv \mathrm{Im}\left\{ \frac{1}{2}\left[ \sum_{nm}{\frac{d^{\omega}}{\hbar}}\frac{d_{nm}^{0}}{\hbar}\varepsilon _{mn}\xi _{mn}^{y}\xi _{nm}^{y}\partial _{k_x}f_{mn}(\vec{k})+\sum_{nm}{\frac{d^{-\omega}}{\hbar}}\frac{d_{nm}^{0}}{\hbar}\varepsilon _{mn}\xi _{mn}^{y}\xi _{nm}^{x}\partial _{k_y}f_{mn}(\vec{k}) \right] \right\}
				\\
				&=\frac{1}{\hbar}\frac{\omega}{\omega ^2+\Gamma ^2}\left\{ \sum_{nm}{\frac{\left( \varepsilon _{nm}/\hbar \right) ^2}{\left( \varepsilon _{nm}/\hbar \right) ^2+\Gamma ^2}\mathrm{Im}\left( \xi _{mn}^{y}\xi _{nm}^{x} \right) \partial _{k_y}f_n(\vec{k})}\right.\\
				&\quad\left.+\sum_{nm}{\frac{\Gamma \varepsilon _{mn}/\hbar}{\left( \varepsilon _{nm}/\hbar \right) ^2+\Gamma ^2}\left[ \xi _{mn}^{y}\xi _{nm}^{y} \partial _{k_x}f_n(\vec{k})-\mathrm{Re}\left( \xi _{mn}^{y}\xi _{nm}^{x} \right) \partial _{k_y}f_n(\vec{k}) \right]} \right\}. \\
		\end{aligned}  \end{equation}
	\end{widetext}

	%%%%%%%%%%%%%%%%%%%%%%%%%%%%%%%%%%%%%%%%%%%%%%%%%%%%%%%%%%%%%%%%%%%%%%%%%%%%%%%%%%%%%%%%%%%%%%%%%%%%%%%%%%%%%%%%%%%%%%%%%%%%%%%%%%%%%%%%%
	%%%%%%%%%%%%%%%%%%%%%%%%%%%%%%%%%%%%%%%%%%%%%%%%%%%%%%%%%%%%%%%%%%%%%%%%%%%%%%%%%%%%%%%%%%%%%%%%%%%%%%%%%%%%%%%%%%%%%%%%%%%%%%%%%%%%%%%%%
	For LPGE (CPGE), considering that the system has time-reversal symmetry, the contribution of the second line of Eq.~(\ref{I_BCD-LPGE-xyy}) [Eq.~(\ref{I_BCD-CPGE-yxy})] to the conductivity can be easily shown to be equal to 0 and we only need to consider the first line of Eq.~(\ref{I_BCD-LPGE-xyy}) [Eq.~(\ref{I_BCD-CPGE-yxy})].
	Only when $\Gamma \ll\left|\varepsilon_{n m}/\hbar\right| $, Eqs.~(\ref{I_BCD-LPGE-xyy}) and (\ref{I_BCD-CPGE-yxy}) can be reduced to the form of the Berry curvature dipole $\sum_n{f_n(\vec{k})}\partial _{k_y}\Omega _{n}^{z} $. Therefore, from Eqs.~(\ref{I_BCD-LPGE-xyy}) and (\ref{I_BCD-CPGE-yxy}) we can see that the BCD term Eq.~(\ref{2nd-conductivity-BCD}) is so complicated that it loses the meaning of ``Berry curvature dipole'' of the traditional BCD formula. From the aspect of PGE conductivities, when $\Gamma \ll\left|\varepsilon_{n m}/\hbar\right| $, using Eq.~(\ref{traditional_BCD_formula}), Eq.~(\ref{2nd-conductivity-BCD}) reduces to
	\begin{equation}
		\begin{aligned}
			&\sigma _{(2)\mathrm{BCD}}^{\eta \alpha \beta}\left( \omega ,-\omega \right)\\
			&\quad \approx -\frac{e^3}{2\hbar ^2}\frac{i}{\omega +i\Gamma}\int{\frac{d^3k}{(2\pi )^3}}\sum_n{\Omega _{n}^{\eta \beta}}\partial _{k_{\alpha}}f_n(\vec{k})\\
			&\quad\quad -\frac{e^3}{2\hbar ^2}\frac{i}{-\omega +i\Gamma}\int{\frac{d^3k}{(2\pi )^3}}\sum_n{\Omega _{n}^{\eta \alpha}}\partial _{k_{\beta}}f_n(\vec{k}).\\
		\end{aligned}
	\end{equation}
	For LPGE,
	\begin{equation} \begin{aligned}  \label{traditional_BCD_formula_new-sigma}
			&\mathrm{Re}\left( \sigma _{(2)\mathrm{BCD}}^{\eta \alpha \beta}\left( \omega ,-\omega \right) \right)\\
			&\quad\approx \frac{e^3}{2\hbar ^2}\frac{\Gamma}{\omega ^2+\Gamma ^2}\int{\frac{d^3k}{(2\pi )^3}}\sum_n{f_n(\vec{k})\left( \partial _{k_{\alpha}}\Omega _{n}^{\eta \beta}+\partial _{k_{\beta}}\Omega _{n}^{\eta \alpha} \right)},
	\end{aligned}  \end{equation}
	which is consistent with Eq.~(5) of Ref.~\onlinecite{Zhang2021} when $\eta \alpha \beta=xyy$ or $ yxx $ \cite{note1/2}. For CPGE,
	\begin{equation} \begin{aligned}
			&\mathrm{Im}\left( \sigma _{(2)\mathrm{BCD}}^{\eta \alpha \beta}\left( \omega ,-\omega \right) \right)\\
			&\quad \approx \frac{e^3}{2\hbar ^2}\frac{\omega}{\omega ^2+\Gamma ^2}\int{\frac{d^3k}{(2\pi )^3}}\sum_n{f_n(\vec{k})\left( \partial _{k_{\alpha}}\Omega _{n}^{\eta \beta}-\partial _{k_{\beta}}\Omega _{n}^{\eta \alpha} \right)}.
	\end{aligned}  \end{equation}
	
	%$\delta_{1,z}=0.031$ eV
	Moreover, we can use numerical results to illustrate the difference between the BCD formula Eq.~(\ref{2nd-conductivity-BCD}) and the traditional BCD formula. At $\tau=5 $ ps, $\nu=1000$ Hz, $\varepsilon_F =0.095$ eV, \textcolor{red}{$E_{\perp}=0.2 \mathrm{~V} / \mathrm{nm}$}, for the NHE or LPGE, using the formulas in Refs.~\onlinecite{Sodemann2015,You2018,Zhang2021} or Eq.~(\ref{traditional_BCD_formula_new-sigma}), we can obtain the traditional Berry curvature dipole and its resulting conductivity $\sigma_{(2) \mathrm{BCD}}^{x y y}(\omega,-\omega)$ of $0.464 ~\text{\AA}$ and $ 4.290 \times 10^4$ $\mathrm{~nm} \cdot \mu \mathrm{A} / \mathrm{V}^2$, respectively. This conductivity result is very close to the result of $4.291 \times 10^4 \mathrm{~nm} \cdot \mu \mathrm{A} / \mathrm{V}^2$ calculated by Eq.~(\ref{2nd-conductivity-BCD}) \cite{note1/2}. In the SM \cite{supplemental_material}, we explain why the results of the two BCD formulas are very close at $\tau=5 $ ps, $\varepsilon_F =0.095$ eV, \textcolor{red}{$E_{\perp}=0.2 \mathrm{~V} / \mathrm{nm}$} and the frequency range of 1000 Hz-1 THz. Hence at this time we can safely use the traditional BCD formula to approximately calculate the LPGE or the DC current of NHE. However, considering the premise of Eq.~(\ref{traditional_BCD_formula}), we set $\tau=10 $ fs, $\nu=1000$ Hz, $\varepsilon_F =0.08$ eV and \textcolor{red}{$E_{\perp}=1 \mathrm{~V} / \mathrm{nm}$}. At this time the traditional Berry curvature dipole and its resulting conductivity are $1.58 ~\text{\AA}$ and $292.15$ $\mathrm{~nm } \cdot \mu \mathrm{A} / \mathrm{V}^2$, respectively. This conductivity result has a large difference of $39.95\mathrm{~nm} \cdot \mu \mathrm{A} / \mathrm{V}^2$ compared to the result of $252.2\mathrm{~nm} \cdot \mu \mathrm{A} / \mathrm{V}^2$ calculated by Eq.~(\ref{2nd-conductivity-BCD}). In addition, we also find that the IB3 term at this time has a contribution of $16.95\mathrm{~nm} \cdot \mu \mathrm{A} / \mathrm{V}^2$. Therefore, it is more accurate to use Eqs.~(\ref{2nd-conductivity-another})-(\ref{2nd-conductivity-IB3-shift}) to calculate the NHE or second-order nonlinear optical responses for some materials with complex energy band structures and femtosecond-scale relaxation times.

	In this work, we used the length-gauge approach to describe the interaction between the electron and the alternating electric field \cite{Passos2018}, while the velocity-gauge approach was used in Refs.~\onlinecite{Xu2021,Passos2018}. The results obtained from the length gauge and the velocity gauge have been shown to be equivalent \cite{Passos2018}. The advantage of length gauge is that conductivities from different terms can be closely linked to insulators or metals \cite{Aversa1995} so that the physics is transparent and neat.
	
	%classified to know which terms disappear in insulators \cite{Aversa1995} and which terms belong to intraband or interband terms.
	
	\textcolor{red}{Some discussion on relaxation time can be seen in the SM \cite{supplemental_material}. Various scattering processes could lead to additional current generation mechanisms \cite{Ganichev2002,Budkin2020,Tarasenko2007,Weber2008} and the effect of these scattering details beyond the relaxation time approximation on the nonlinear optical responses of WTe$_2$ monolayer deserves future exploration.} Besides, we do not consider the electron-electron interaction and its resulting exciton effect in the present work. However, both theories and experiments show that the exciton effect has a significant impact on the optical response of 2D materials \cite{Xie2019,Mkrtchian2019,Zhumagulov2022}. How to accurately and completely include the complex exciton effect on PGE and SHG in the quantum kinetics method still deserves further study.	
	
	\textcolor{red}{The quantum kinetic method can also be extended for the study of transport phenomena under magnetic fields. For example, in Ref. \onlinecite{Cullen2021}, the anomalous planar Hall effect caused by applying an in-plane magnetic field in a 2D heavy-hole system was found to be a purely intrinsic phenomenon, and this novel effect can serve as an ingenious scheme to unambiguously probe the effect of Berry curvature on transport; Bhalla $et ~al.$ \cite{Bhalla2020} proposed that applying linearly polarized light on the surface of a doped topological insulator with in-plane magnetization would lead to a second-order DC current with resonant feature, and this new effect is called resonant photovoltaic effect. The influence of applying a magnetic field or performing exchange bias on the nonlinear optical responses of WTe$_2$ monolayer remains a subject of future study. In addition, a quantum kinetic theory for the nonlinear response of ballistic topological edge states has been developed \cite{Bhalla2021}, so the nonlinear optical response of the helical edge states of the quantum spin Hall insulator WTe$_2$ monolayer in various frequency regions is also worthy of future exploration.}
	
	%%%%%%%%%%%%%%%%%%%%%%%%%%%%%%%%%%%%%%%%%%%%%%%%%%%%%%%%%%%%%%%%%%%%%%%%%%%%%%%%%%%%%%%%%%%%%%%%%%%%%%%%%%%%%%%%%%%%%%%%%%%%%%%%%%%%%%%%%
	  %Use the figure* environment to get a wide figure that spans the page in \texttt{twocolumn} formatting
	%%%%%%%%%%%%%%%%%%%%%%%%%%%%%%%%%%%%%%%%%%%%%%%%%%%%%%%%%%%%%%%%%%%%%%%%%%%%%%%%%%%%%%%%%%%%%%%%%%%%%%%%%%%%%%%%%%%%%%%%%%%%%%%%%%%%%%%%
	%%%%%%%%%%%%%%%%%%%%%%%%%%%%%%%%%%%%%%%%%%%%%%%%%%%%%%%%%%%%%%%%%%%%%%%%%%%%%%%%%%%%%%%%%%%%%%%%%%%%%%%%%%%%%%%%%%%%%%%%%%%%%%%%%%%%%%%%%
	  %Use the figure* environment to get a wide figure that spans the page in \texttt{twocolumn} formatting
	%%%%%%%%%%%%%%%%%%%%%%%%%%%%%%%%%%%%%%%%%%%%%%%%%%%%%%%%%%%%%%%%%%%%%%%%%%%%%%%%%%%%%%%%%%%%%%%%%%%%%%%%%%%%%%%%%%%%%%%%%%%%%%%%%%%%%%%%%

	%%%%%%%%%%%%%%%%%%%%%%%%%%%%%%%%%%%%%%%%%%%%%%%%%%%%%%%%%%%%%%%%%%%%%%%%%%%%%%%%%%%%%%%%%%%%%%%%%%%%%%%%%%%%%%%%%%%%%%%%%%%%%%%%%%%%%%%%%
	  %Use the figure* environment to get a wide figure that spans the page in \texttt{twocolumn} formatting
	%%%%%%%%%%%%%%%%%%%%%%%%%%%%%%%%%%%%%%%%%%%%%%%%%%%%%%%%%%%%%%%%%%%%%%%%%%%%%%%%%%%%%%%%%%%%%%%%%%%%%%%%%%%%%%%%%%%%%%%%%%%%%%%%%%%%%%%%%
	
	  %Use the figure* environment to get a wide figure that spans the page in \texttt{twocolumn} formatting
	
	\section{Acknowledgements}
	We thank Yuan-Dong Wang and Zhi-Fan Zhang for useful discussions. This work is supported in part by the NSFC (Grants No. 11974348, and No. 11834014), and the National Key R\&D Program of China (Grant No. 2018YFA0305800, No. 2022YFA1402800), and the Strategic Priority Research Program of CAS (Grants No. XDB28000000 and No. XDB33000000). ZGZ is supported in part by the Training Program of Major Research plan of the National Natural Science Foundation of China (Grant No. 92165105), and CAS Project for Young Scientists in Basic ResearchGrant No. YSBR-057.

%	\appendix{}
%	\setcounter{figure}{0}
%	\renewcommand{\thefigure}{S\arabic{figure}}
%	
%	
	\bibliography{ref}

\end{document}